\documentclass[acmsmall, screen]{acmart}

\usepackage{float}

\usepackage{booktabs}   
\usepackage{subcaption} 

\usepackage{todonotes}
\usepackage{xspace}
\usepackage{enumitem}
\usepackage{framed}
\usepackage[most]{tcolorbox}

\usepackage{todonotes}
\usepackage{amsmath,amsfonts}
\usepackage{algorithmic}
\usepackage{graphicx}
\usepackage{xcolor}
\usepackage{svg}
\def\BibTeX{{\rm B\kern-.05em{\sc i\kern-.025em b}\kern-.08em
    T\kern-.1667em\lower.7ex\hbox{E}\kern-.125emX}}

\usepackage{listings}%
\newcommand{\li}[1]{\lstinline{#1}}

\usepackage{color}

\lstdefinestyle{ocamlboxed}{
  language={caml},
  tabsize=2, 
  basicstyle=\ttfamily\small,
  commentstyle=\itshape\ttfamily\color{gray}, 
  numbers=none,
  morekeywords={module,struct,sig,val,ref,bool, case},
  showspaces=false,
  showstringspaces=false,
  stringstyle=\ttfamily\color{olive},
  mathescape=true,
  escapechar=\#,
  frame=single,
  linewidth=.99\linewidth,
  xleftmargin=0.1cm,
}
\lstdefinestyle{haskellpseudo}{
  language={},  
  morekeywords={let, in, where, case, of, class, data, instance, type, if, then, else, do, import, module},
  sensitive=true,
  commentstyle=\color{gray}\itshape,
  keywordstyle=\color{black}\bfseries,
  stringstyle=\color{red},
  morestring=[b]",
  morecomment=[l]--,
  morecomment=[n]{\{-}{-\}},
  literate={->}{$\rightarrow$}1 {<-}{$\leftarrow$}1 {=>}{$\Rightarrow$}1 {::}{$:$}1 {forall}{$\forall$}1,
  basicstyle=\ttfamily\small,
  breaklines=true,
  frame=single,
  numbers=none,
  numberstyle=\tiny,
  captionpos=b,
  tabsize=2
}

\definecolor{highlighteryellow}{RGB}{250,220,120}
\definecolor{typeTyrian}{RGB}{129,76,245}
\definecolor{patBlue}{RGB}{23,160,221}

\newcommand{\codetype}[1]{\textcolor{typeTyrian}{\li{#1}}}
\newcommand{\codepat}[1]{\textcolor{patBlue}{\li{#1}}}

\newcommand{\highlightli}[1]{%
  {%
    \setlength{\fboxsep}{1.5pt}
    \colorbox{highlighteryellow}{\li{#1}}%
  }%
}

\setcopyright{rightsretained}
\acmDOI{10.1145/3689728}
\acmYear{2024}
\acmJournal{PACMPL}
\acmVolume{8}
\acmNumber{OOPSLA2}
\acmArticle{288}
\acmMonth{10}
\acmSubmissionID{oopslab24main-p278-p}
\received{2024-04-06}
\received[accepted]{2024-08-18}

\begin{document}

\title{Statically Contextualizing Large Language Models with Typed Holes}

\author{Andrew Blinn}
\orcid{0000-0001-6938-7379}
\affiliation{%
  \institution{University of Michigan}
  \city{Ann Arbor}
  \country{USA}
}
\email{blinnand@umich.edu}

\author{Xiang Li}
\orcid{0009-0005-6860-039X}
\affiliation{%
  \institution{University of Michigan}
  \city{Ann Arbor}
  \country{USA}
}
\email{xkevli@umich.edu}

\author{June Hyung Kim}
\orcid{0009-0005-0820-9532}
\affiliation{%
  \institution{University of Michigan}
  \city{Ann Arbor}
  \country{USA}
}
\email{jpoly@umich.edu}

\author{Cyrus Omar}
\orcid{0000-0003-4502-7971}
\affiliation{%
  \institution{University of Michigan}
  \city{Ann Arbor}
  \country{USA}
}
\email{comar@umich.edu}


\begin{abstract}
Large language models (LLMs) have reshaped the landscape of program synthesis. 
However, contemporary LLM-based code completion systems often hallucinate broken 
code because they lack appropriate code context, particularly when working with definitions 
that are neither in the training data nor near the cursor. 
This paper demonstrates that tighter integration with the type and binding structure of the programming 
language in use, as exposed by its language server, can help address this contextualization problem in a token-efficient manner.
In short, we contend that AIs need IDEs, too! 
In particular, we integrate LLM code generation into the Hazel live program sketching environment.
The Hazel Language Server is able to identify the type and typing context of the hole that the programmer is filling, 
with Hazel's total syntax and type error correction ensuring that a meaningful program sketch is available whenever the developer requests a completion. 
This allows the system to prompt the LLM with codebase-wide contextual information that is not lexically local to the cursor, nor necessarily in the same file, but that is likely to be semantically local to the developer's goal. Completions synthesized by the LLM are then iteratively refined via further dialog with the language server, which provides error localization and error messages. To evaluate these techniques, we introduce MVUBench, a dataset of model-view-update (MVU) web applications with accompanying unit tests that have been written from scratch to avoid data contamination, and that can easily be ported to new languages because they do not have large external library dependencies. These applications serve as challenge problems due to their extensive reliance on application-specific data structures.  
Through an ablation study, we examine the impact of contextualization with type definitions, function headers, and errors messages, individually and in combination. We find that contextualization with type definitions is  particularly impactful. 
After introducing our ideas in the context of Hazel, a low-resource language, we duplicate our techniques and port MVUBench to TypeScript in order to validate the applicability of these methods to higher-resource mainstream languages.
Finally, we outline ChatLSP, a conservative extension to the Language Server Protocol (LSP) that language servers can implement to expose capabilities that AI code completion systems of various designs can use to incorporate static context when generating prompts for an LLM.
\end{abstract}

\ccsdesc[500]{Software and its engineering~Software creation and management}
\ccsdesc[500]{Theory of computation~Type structures}

\keywords{Large Language Models, Program Synthesis, Program Repair}

\maketitle

\begin{figure}[h]
  \includegraphics[width=\linewidth]{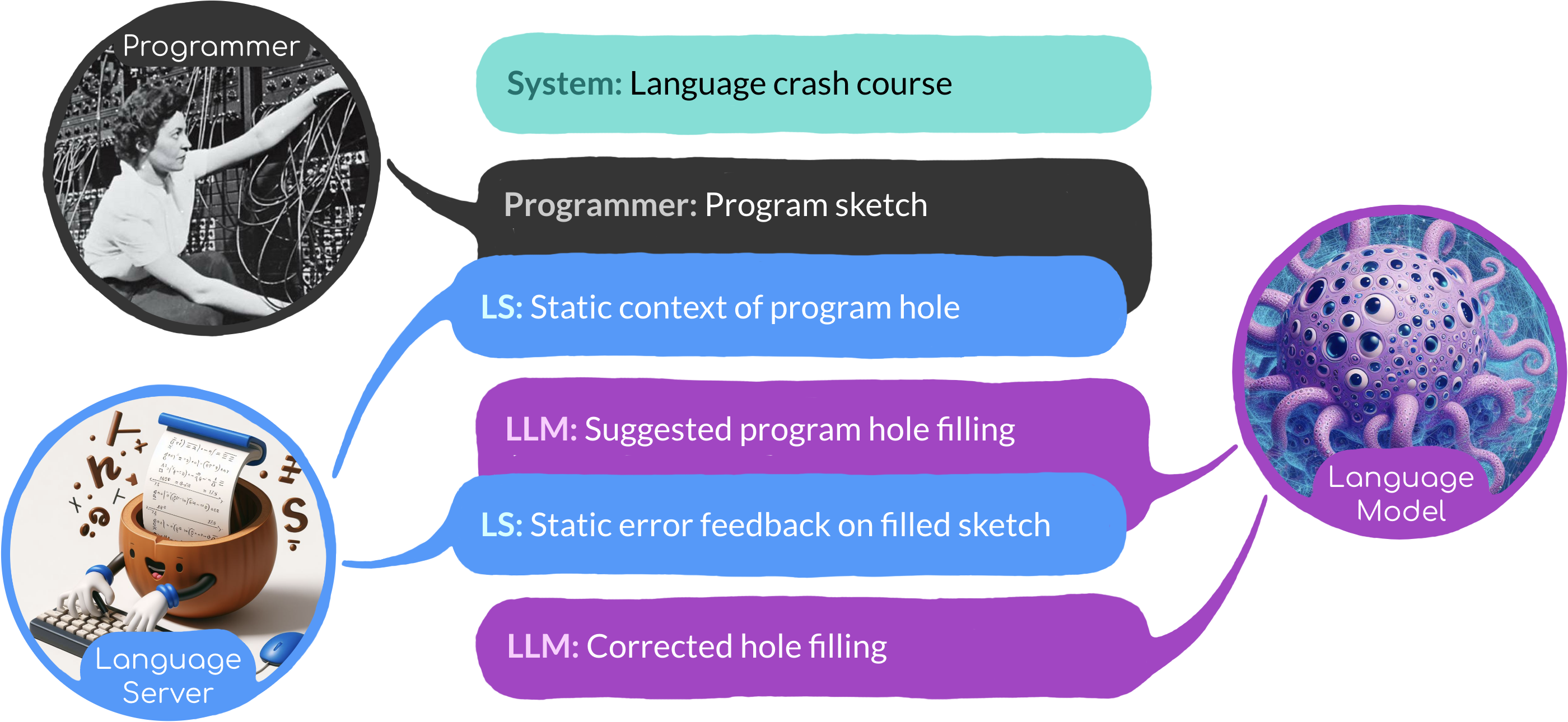}
  \caption{Hazel Assistant Conversational Architecture (see \autoref{sec:acknowledgements} for image acknowledgements)}
  \label{fig:conv-arch}
\end{figure}

\section{Introduction}
\label{sec:introduction}

Recent advances in generative AI have triggered an avalanche of new AI programming assistants---the most prominent is Copilot~\cite{Copilot}, but it has many competitors---that generate code completions by prompting a large language model (LLM) pre-trained on a corpus of diverse natural language documents as well as code written in various programming languages~\cite{DBLP:journals/corr/abs-2108-07732,chen2021evaluating,DBLP:conf/icse/VaithilingamGGGHMPRSWY23,DBLP:conf/pldi/Xu0NH22,chen2021evaluating,embeddings}. 
Once trained, an LLM iteratively transforms an input token sequence, called the \emph{prompt}, into \emph{next-token probability distributions} from which  \emph{completions} are sampled. 
LLMs are able to learn statistical regularities in the training data~\cite{hindle2016naturalness}, with limited reasoning abilities emerging as LLMs scale up in size \cite{Emergent2022}. As a result, AI assistants have become capable enough to substantially impact developer productivity \cite{DBLP:journals/cacm/ZieglerKLRRSSA24,DBLP:journals/corr/abs-2302-06590,DBLP:conf/chi/Vaithilingam0G22,DBLP:journals/pacmpl/BarkeJP23}. 
For example, one study  reports a 50\% increase in productivity when using Copilot \cite{DBLP:journals/cacm/ZieglerKLRRSSA24}. 
The impact is particularly pronounced for developers working with  \emph{high resource} libraries and languages, i.e. those well-represented in the training data~\cite{DBLP:conf/pldi/Xu0NH22}.




Contemporary AI assistants construct the prompt primarily using the program text appearing in a textual window around the developer's cursor (the \emph{cursor window})~\cite{semenkin2024context}. This approach leads to poor performance in situations where critical task-relevant context comes from definitions that appear neither in the cursor window nor in the training data (the \textbf{semantic contextualization problem})~\cite{zhang2023repocoder,CopilotingTheCopilots,ShrivastavaRepo,pei2023better,li2024enhancing,ding2023cocomic,agrawal2023guiding,zan2023private}. For example, consider the following cursor window, which would arise when a developer is implementing a GUI component using the model-view-update (MVU) architecture (central to popular GUI application frameworks like React~\cite{MVUReact} and Elm~\cite{ElmArchitecture}): 

\begin{lstlisting}[style=ocamlboxed]
  (* update the room booking data after a user action *)
  function update(model: Model, action: Action): Model {
\end{lstlisting}

Correctly completing this function definition requires knowing the definitions of this specific component's  \li{Model} and \li{Action} types, which commonly appear in different files in the repository and therefore outside the cursor window. Various other files might also contain relevant definitions, e.g. other types that can be reached from the definitions of \li{Model} and \li{Action}, and useful helper functions for working with values of these types. Without access to these definitions, an LLM will either be unable to generate sufficiently probable completions (which may result in no completion being generated) or more typically it will hallucinate plausible-but-incorrect definitions based only on the provided comment~\cite{zhang2023siren,xu2024hallucination}. 

To address this problem, assistant designers use various retrieval augmented generation (RAG) techniques \cite{DBLP:journals/corr/abs-2312-10997} to retrieve additional code from other files in the repository and external libraries for inclusion in the prompt. Real-world code bases often span hundreds of thousands of lines of code, so exhaustive retrieval quickly runs into scaling issues. While prompt (i.e. context) size limits continue to increase~\cite{ding2024longrope}, generation costs (measured in both time and energy) scale with token count~\cite{hoffmann2022training}. These costs are substantial (because LLMs typically have billions of parameters) so \emph{token efficiency} remains a critical metric~\cite{semenkin2024context}. Moreover, contemporary LLMs struggle to attend to relevant information and ignore irrelevant information (such as the \li{Model} and \li{Action} types for \emph{other} GUI components) in large prompts~\cite{xu2024knowledge,levy2024same,li2023loogle,liu2024lost}.

Given these issues, assistant designers need retrieval techniques that prioritize task-relevant code. For example, Copilot retrieves code from locations in other files that the developer has recently visited, based on the heuristic that these are likely to be task-relevant~\cite{Thakkar_2022}. Another prominent retrieval strategy, which we will refer to as \emph{vector retrieval}, involves performing a vector search across the repository (and perhaps beyond) to retrieve code similar to the code in the cursor window~\cite{zhang2023repocoder,zan2023private,lu2022reacc}, as measured by a learned vector similarity metric~\cite{embeddings,lewis2021retrievalaugmented}. This relies on the heuristic that lexically similar code is likely to be task-relevant code. In the example above, since the type names \li{Model} and \li{Action} appear explicitly in the cursor window, this approach \emph{may} find their definitions. However, it may also find irrelevant definitions of other types named \li{Model} and \li{Action} and other implementations of \li{update}, e.g. those from other GUI components in this or other applications. It may also be less effective when the task-relevant definitions are not named explicitly, e.g. if the developer is later writing a {call} to \li{update}, the fact that the relevant types are \li{Model} and \li{Action} requires reasoning about the type signature of the \li{update} function.  

These retrieval approaches are language-agnostic, treating source code as  a sequence of tokens like any other, so they must necessarily deploy imprecise heuristics. In this paper, we instead consider language-aware approaches that take advantage of the fact that in many languages, code is governed by a rich type and binding discipline determined by a static semantics~\cite{tapl,pfpl}. 
To retrieve relevant semantic information and analyze candidate code completions, we rely on a modern language server~\cite{lsp-study,bour2018merlin} to provide various \emph{language services}, namely type reporting, typing context search, and error reporting. Integrated development environments (IDEs) interact with these services to drive various human-facing affordances such as type hints and hover messages. Here we investigate the hypothesis that LLM code completion can also benefit from interactions with language services. Put pithily, we hypothesize that \emph{AIs need IDEs, too}.

We investigate two language-aware approaches independently and in combination:

\paragraph{Static Retrieval}
The first approach we consider (\autoref{sec:semantic-retrieval}) is static retrieval, where the language server is tasked to (1) determine the type and typing context of the ``hole'' (implicit in the \li{update} sketch above) at the cursor, and (2) transitively retrieve semantically relevant type definitions and function headers, from wherever they might appear, for inclusion in the prompt.  

In the \li{update} example, the hole that the language server inserts (either implicitly or explicitly) at the cursor is of type \li{Model} (because it is in the body of the \li{update} function, which has return type \li{Model}), and the local typing context includes the argument \li{action : Action}, so the language server can  look up the definitions of the \li{Model} and \li{Action} types. These types might themselves refer to other types, so we can transitively continue type retrieval. We can also retrieve information about relevant helper functions in the typing context, e.g. those that operate on the types that have been looked up, continuing transitively up to a token limit.


\paragraph{Static Error Correction}
To further improve correctness, we combine static retrieval in \autoref{sec:error-correction} with a straightforward {syntactic and static error correction} pass \cite{joshi2022repair,STALL}: we ask the language server to localize and generate error messages for syntactic and static errors in the generated completion, then feed this information back into an instruction-tuned model, prompting it to correct these errors, potentially over multiple rounds (trading off latency for correctness).


\subsection{Evaluation Overview}

\subsubsection{Programming Languages and Language Servers}
The approaches that we investigate in this paper are in principle applicable to any programming language with a well-structured type and binding discipline. The more challenging requirement, which has limited the prior experiments in these directions (which we review in \autoref{sec:related-work}), is that we also need a rather capable language server. In particular, the language server must be capable of robust syntax error and type error recovery, producing a semantically meaningful program sketch (i.e. a program with \emph{holes}) in any situation where the developer might request code completion~\cite{HazelnutSNAPL}. For instance, the example situation from the beginning of \autoref{sec:introduction} are syntactically erroneous, because the developer has not yet closed a curly brace or parenthesis, so a standard compiler would simply report a syntax error and halt. In other cases, there may be localized type errors elsewhere in the program that would ideally not cause gaps in the availability of code completion. 

For this reason, we evaluate these ideas primarily by developing an AI programming assistant for Hazel, extending the Hazel Assistant with LLM support~\cite{vlhcc2022assistant}. Hazel is a typed functional programming environment designed specifically around typed holes, inserting them automatically to ensure total syntax error recovery~\cite{tylr2023} and total type error recovery~\cite{hazel-popl24}. 
Hazel is also capable of evaluating programs with holes~\cite{HazelnutLive,peanut-oopsla23} (including `non-empty' holes inserted as membranes around marked type errors~\cite{hazel-popl24}), which makes it possible to use unit testing to granularly evaluate the correctness of even locally ill-typed model outputs (rather than the more ad hoc methods that are common in the literature, like edit distance from a single canonical solution). 

The Hazel language is similar to Elm, OCaml, and other languages in the ML family, i.e. it is a pure typed functional language with support for algebraic datatypes and pattern matching. Unlike Elm and OCaml, contemporary LLMs have not been trained on a substantial body of Hazel code, i.e. Hazel is a \emph{low resource language}. 
This presents both a challenge and an opportunity for research. 
We have found that when asked to write Hazel code, contemporary LLMs fail to follow Hazel's syntax and semantics, often borrowing syntactic forms and library functions from OCaml and Elm. However, LLMs are capable of in-context learning~\cite{Gao_2023,radford2019language}, suggesting that it is possible to include few-shot examples and instructions in the prompt to quickly teach contemporary LLMs about how Hazel differs from related higher resource languages. 
The error correction approach we investigate may also be of particular interest in preventing errors in this sort of low-resource setting, which is of considerable interest to the PL research community~\cite{cassano2024knowledge}. 

To demonstrate that static retrieval is useful even for mainstream high resource languages, we also perform additional more limited experiments in \autoref{sec:retrieval-typescript} with TypeScript via the TypeScript Language Server, an instance of the Language Sever Protocol (LSP)~\cite{lsp-study}. We find that the LSP does not provide simple, direct access to the sort of information that is necessary to implement static retrieval, so we propose a more direct interface as we introduce the various approaches throughout the paper, summarized as a prospective LSP extenstion in \autoref{sec:chatlsp}.


\subsubsection{Tasks}
\label{sec:tasks}
The most commonly reported LLM code completion benchmarks are HumanEval~\cite{chen2021evaluating}, EvalPlus~\cite{liu2023code}, MBPP~\cite{DBLP:journals/corr/abs-2108-07732}, and LiveCodeBench~\cite{DBLP:journals/corr/abs-2403-07974}. 
These are unsuitable for evaluating the proposed approaches because they consist of single-file tasks (constructed either manually, or derived from public repositories or programming contests) that are low-context, i.e. they require only builtin or standard datatypes, and therefore do not highlight the semantic contextualization problem.


\emph{Repository-level} benchmarks like RepoEval~\cite{zhang2023repocoder}, RepoBench~\cite{DBLP:journals/corr/abs-2306-03091}, CrossCodeEval~\cite{ding2023crosscodeeval}, the CoCoMIC dataset~\cite{ding2023cocomic}, and defects4j~\cite{DBLP:conf/issta/JustJE14} are more suitable because they are high-context, i.e. they require completing code that depends on definitions in different files. However, these benchmarks also present difficulties:
\begin{enumerate}
\item \textbf{Data Contamination}: LLMs are known to be able to memorize code that they have seen during training, and evidence suggests that this data contamination issue has indeed caused public models to be overfitted to publicly available code \cite{DBLP:journals/corr/abs-2403-07974,sainz2023nlp}. All of these benchmarks source examples from GitHub or PyPI. Based on reported cut-off dates, the projects in these benchmarks will have by now likely been incorporated into the training of contemporary models (noting that with few exceptions discussed below, training data is not disclosed).

\item \textbf{Language Exclusivity}: None of these include Hazel code, nor is it easy to manually port arbitrary projects taken from GitHub or PyPI that depend on various complex libraries to Hazel or any other low resource language of interest to the community. Existing porting techniques have only generated low-context datasets~\cite{cassano2024knowledge}. In addition, RepoEval and CoCoMIC exclusively feature Python code, which is difficult to statically analyze.

\item \textbf{Missing Tests}: None of the tasks in these benchmarks include unit tests, which means that we can only evaluate correctness based on brittle textual similarity metrics. We observe that models often produce correct output with substantial syntactic variation, similar to human programmers \cite{glassman2015}.
\end{enumerate}


For these reasons, we construct a repository-level benchmark suite, MVUBench, in \autoref{sec:mvubench}, that consists of various MVU web applications.
Web application development is an important and under-studied application domain. Many visions for the future imagine LLMs generating complete application logic, not just solving code competition problems \cite{elena2022}.
MVU applications are high-context in that they typically define a number of different datatypes which by convention are often located in separate files. Some of these datatypes have generic names, like \li{Model} and \li{Action}, and a single application might have multiple such types, one for each GUI component,  
presenting a significant challenge to language-agnostic techniques that do not understand binding structure. Indeed, it is easy to construct particularly challenging yet realistic examples simply by combining multiple components of MVUBench, as we demonstrate in \autoref{sec:vector-RAG}.

We address the data contamination problem following the approach taken by HumanEval, by conceptualizing and implementing these applications from scratch, without directly adapting any code from GitHub. We will control the release of these benchmarks to limit the likelihood of future contamination~\cite{jacovi-etal-2023-stop}. We address the language exclusivity problem by ensuring that these MVU applications do not have any external library dependencies, so it is easy to port them to languages beyond Hazel and TypeScript, notably including pure functional languages. New MVU examples are also easy to develop and add to the benchmark, because they can implement the logic of essentially any front-end web application or GUI component. Finally, the lack of side effects also makes it easy to unit test the core application logic.

\subsubsection{LLMs}
We selected two pre-trained language models with which to perform experiments. First, we selected OpenAI's GPT-4(-0613)~\cite{achiam2023gpt}, which is currently consistently at or near the top of code completion benchmarks, to evaluate whether even the most capable contemporary foundation models (i.e. models so large that only large organizations like OpenAI have the resources to train and deploy them) benefit from the approaches we consider. 

GPT-4 is a closed model and many of its specific details, including its size and training, have not been publicly disclosed. This presents a significant challenge to reproducibility. Consequently, we also conduct experiments using StarCoder2-15B, the most capable fully open-source model responsibly trained on a fully open code corpus, The Stack v2, as of this writing~\cite{lozhkov2024starcoder}. The Stack v2 has notably extensive coverage of low resource languages. StarCoder2-15B is small enough to run locally on sufficiently powerful workstations, so our results should be reproducible
indefinitely.

GPT-4 is an instruction-tuned model, so we are able to use it to evaluate static error correction. StarCoder2-15B is completion-tuned, meaning it is not designed to receive and respond to instructions. We are not aware of a comparably powerful fully open source code-specialize model that is instruction-tuned as of this writing, so we do not evaluate error correction with StarCoder2-15B.

\section{Static Retrieval and Error Correction in the Hazel Assistant}
\label{sec:semantic-retrieval}

We first introduce Hazel and the Hazel Assistant by example from the developer's perspective. We continue by describing how the Hazel Assistant prompts GPT-4 
and interfaces with the Hazel Language Server to generate code completions augmented with static retrieval and error correction. Then, we introduce the MVUBench benchmark suite and report the results of an ablation study of each of these features. We investigate their relative contributions to the overall performance of the assistant on these high-context MVU tasks, relative to various baselines that establish bounds on performance.

\subsection{Hazel}
\label{sec:hazel}
\label{sec:hazel-assistant}

\noindent
Hazel is a web-based live functional programming environment that features total syntax and type error recovery via automatic hole insertion~\cite{tylr2023,hazel-popl24}. This ensures that every editor state in Hazel is a semantically meaningful program sketch and that Hazel's various editor services, include code completion, never experience gaps in service~\cite{HazelnutSNAPL}. 

\begin{figure}[h]
  \includegraphics[width=13.5cm]{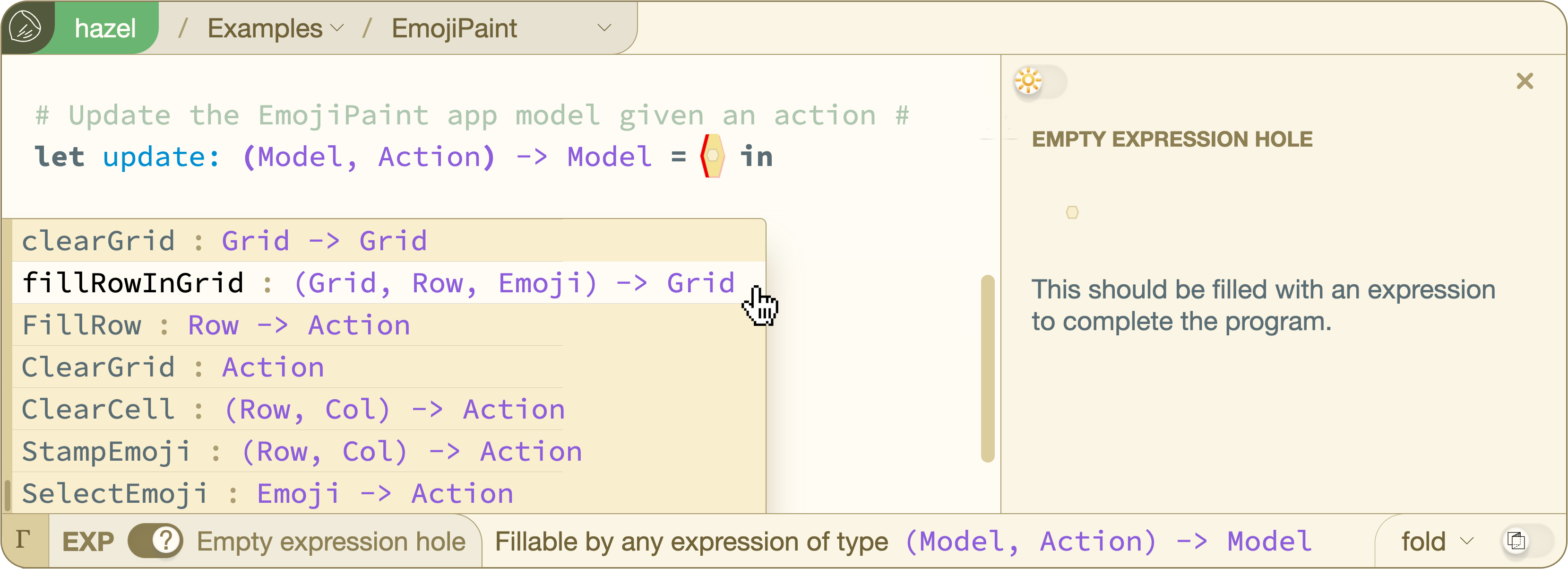}
  \caption{The Hazel IDE}
  \label{fig:hazel-editor-1}
\end{figure}

As a running example, consider a scenario where the developer is implementing EmojiPaint, a simplified MVU app where a user chooses an emoji from a palette and paints designs by `stamping' it on a grid. \autoref{fig:hazel-editor-1} shows the user editing the file where the \li{update} function is defined. 
In \autoref{fig:hazel-editor-1} the developer's cursor is shown as a red convex triangle to the right of a hole, represented as a convex hexagon. The Hazel IDE interfaces with the Hazel Language Server to report static information which both user and model can use to inform completions. The bottom bar, called the Cursor Inspector, reports information on the syntax as well the expected type of the expression, here a typed hole of the function type \codetype{(Model, Action) -> Model}, at the cursor. The lower left popup, called the Context Inspector, reports the typing context at the cursor.

\begin{figure}[h]
  \includegraphics[width=\linewidth]{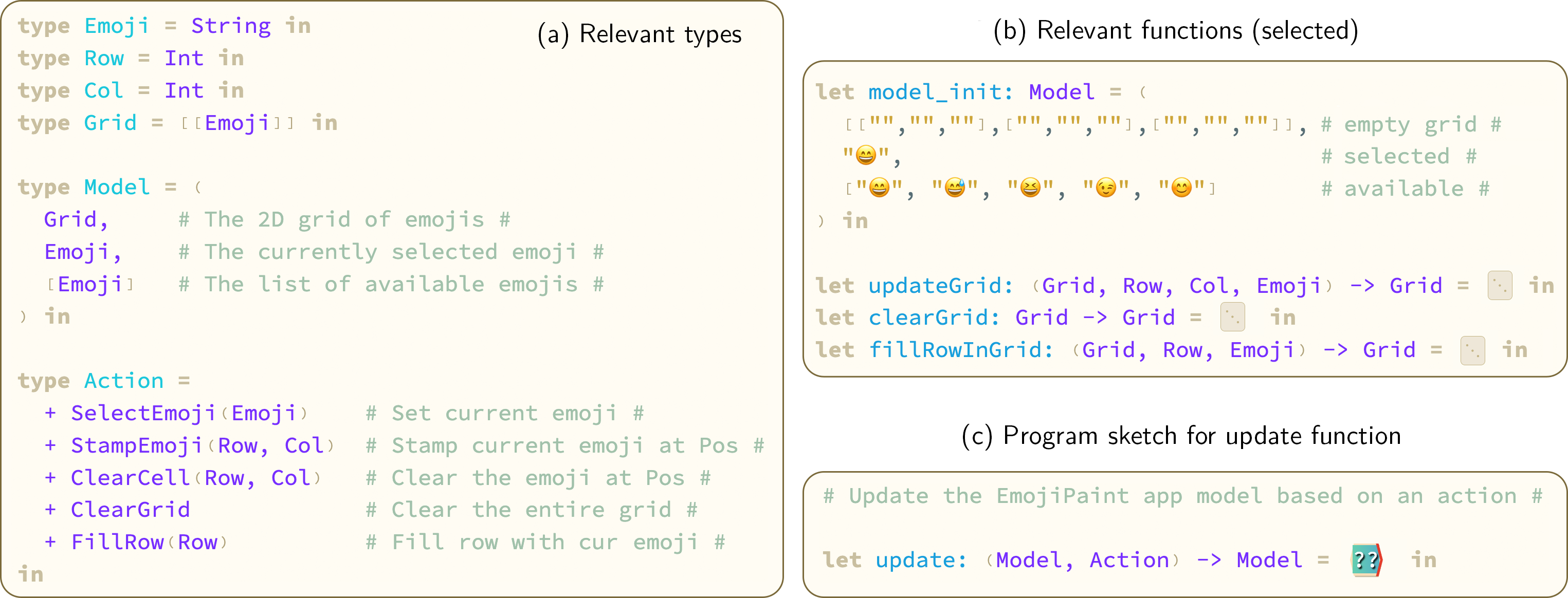}
  \caption{(a) The types relevant to the EmojiPainter MVU app. (b) An excerpt of already-implemented functions in another file (with definitions collapsed by Hazel's outliner). (c) The stubbed function header, where the developer has requested LLM completion by inserting \highlightli{??} in the hole.}
  \label{fig:todo-whole-sketch}
\end{figure}

The \codepat{update} function is intended to respond to EmojiPaint user actions in the GUI, represented by values of type \codetype{Action}, transforming the current GUI state, represented by values of type \codetype{Model}, to a new GUI state, also of type \codetype{Model}. These types and associated helper functions appear in different files, excerpted in \autoref{fig:todo-whole-sketch}(a-b). Algebraic datatypes (i.e. recursive sum types) are represented in Hazel as constructors separated by \codetype{+} (rather than \codetype{|} in similar languages like OCaml).

\subsection{Hazel Assistant}
\begin{figure}[h]
  \includegraphics[width=11.5cm]{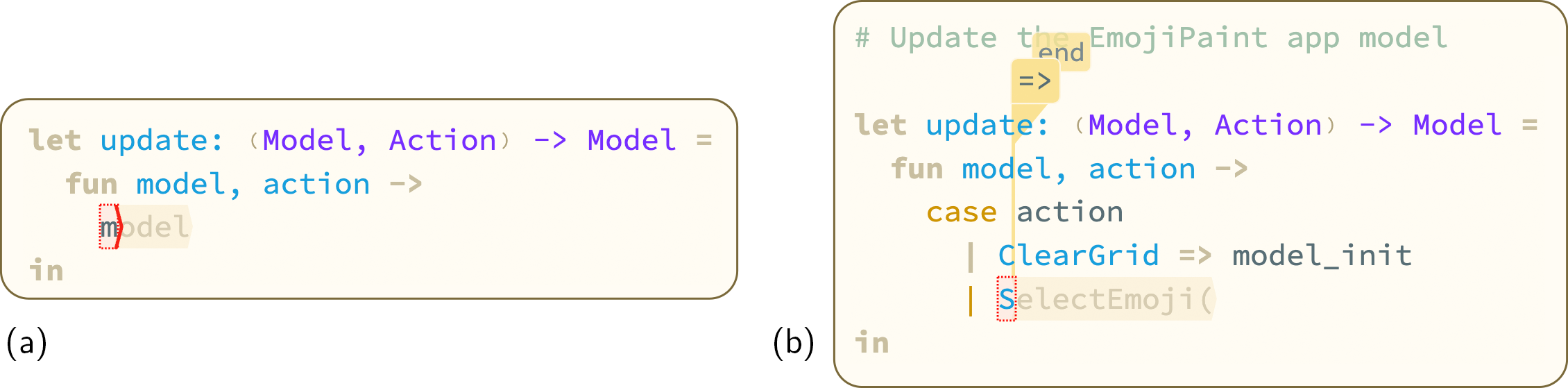}
  \caption{The Hazel Assistant defaults to providing only type-directed token completion.}
  \label{fig:hazel-assistant-base}
\end{figure}

The Hazel Assistant is a programming assistant that generates code completions by two mechanisms. To provide fast, local completions, the Hazel Assistant generates type-directed completions~\cite{vlhcc2022assistant}, using localized syntactic and static information to inform small completions with type-directed lookahead as shown in \autoref{fig:hazel-assistant-base}(a-b). 
This feature can be invoked even when there are syntax errors because 
Hazel tracks syntactic obligations in a backpack, e.g. as shown in \autoref{fig:hazel-assistant-base}(b) where both \highlightli{=>} and \highlightli{end} are necessary to complete the case expression~\cite{tylr2023}.

\begin{figure}[h]
  \includegraphics[width=12cm]{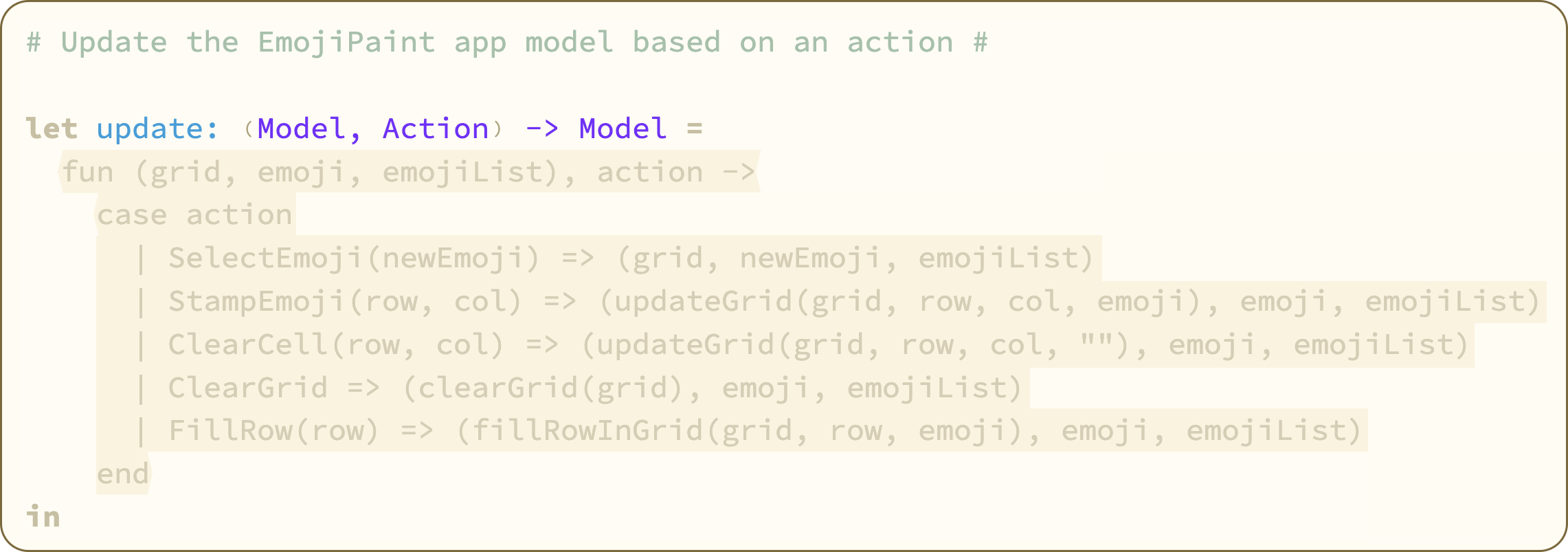}
  \caption{The Hazel LLM Assistant, combining static information with generative creation via language models, is capable of offering more substantial completions}
  \label{fig:hazel-llm-assistant}
\end{figure}

To request an LLM completion from the Hazel Assistant, the developer can fill any expression hole with \highlightli{??} which starts to animate as suggested in \autoref{fig:todo-whole-sketch}(c). GPT-4, our underlying model in this section, is not particularly fast as of this writing, so the developer can continue to edit elsewhere while waiting for GPT-4 to return a completion. For this example, \autoref{fig:hazel-llm-assistant} shows an example of a GPT-4-generated completion. The developer can inspect this completion (which would display any type errors found) and accept it with the \li{Tab} key.

\subsection{The Hazel Assistant Trialogue}

Our generative hole filling process consists of the following steps, construed here as a `trialogue' between programmer, the Hazel Language Server, and a language model. This is depicted in \autoref{fig:conv-arch} as a series of chat messages. In our setup, the primary interaction is between the language model and the Hazel Language Server, which acts on behalf of the user in response to a request for LLM hole filling, kicking off the following exchange. Here we use the messaging terminology from the OpenAI Chat API \cite{openAIChatAPI}, which distinguishes System, User, and Model messages: 
\begin{enumerate}
    \item \emph{System Message}: Hazel Crash Course and few-shot examples 
    \item \emph{User Message}: Program sketch augmented by static retrieval\\ (i.e. relevant semantic context from the language server)
    \item \emph{Model Message}: Suggested hole filling
    \item \emph{User Message}: Syntax and type errors in the completion, if any
    \item \emph{Model Message}: A corrected completion, if necessary
\end{enumerate}

We repeat steps 4-5, i.e. we perform syntactic and static error correction when needed, stopping after at most two iterations to limit latency.

\subsection{System Message: The Hazel Crash Course}

The system message is generic, common to each prompt. For an instruction-tuned model (GPT-4), the system message consists of three sections:

First, we provide a list of instructions delineating the task. In particular, we instruct the model to provide a code fragment to replace a sentinel value representing the target hole in the program sketch. For example, the model is given the instructions:
\begin{itemize}
    \item \li{"Reply only with code"}
    \item \li{"DO NOT include the program sketch in your reply"}
\end{itemize}

Second, an informal specification of Hazel syntax with emphasis on `negative characterization', listing differences from syntactically-similar higher-resource languages. As this kind of `prompt engineering' is as-yet a task-sensitive and inexact process, this section, along with the one above, was constructed though an ad-hoc process of discovering repeated syntactic errors in model output. For example:

\begin{itemize}
        \item \li{"No 'rec' keyword is necessary for 'let' to define a recursive function"}
        \item \li{"There is no dot accessor notation for tuples; use pattern matching"}
\end{itemize}

Finally, we positively characterise hazel syntax via a fixed list of input/output pairs of sketches and program completions (few-shot prompting). We show one example below:

\begin{itemize}
        \item \textbf{Sketch} \begin{lstlisting}[style=ocamlboxed]
let List.length: [(String, Bool)] -> Int =
  fun xs -> ?? end in \end{lstlisting}
        \item \textbf{Completion} \begin{lstlisting}[style=ocamlboxed]
case xs
| [] => 0
| _::xs => 1 + List.length(xs)
end
        \end{lstlisting}
\end{itemize}

For the smaller completion model (StarCoder2-15B) discussed below, which has a longer context window (16k versus 8k for GPT-4-0613) but is not instruction-tuned, we omit the first two sections, in lieu of providing a longer list of syntax examples, which are provided simply as a list of definitions rather than input-output pairs.

\subsection{Type Retrieval}

\begin{figure}[h]
  \includegraphics[width=\linewidth]{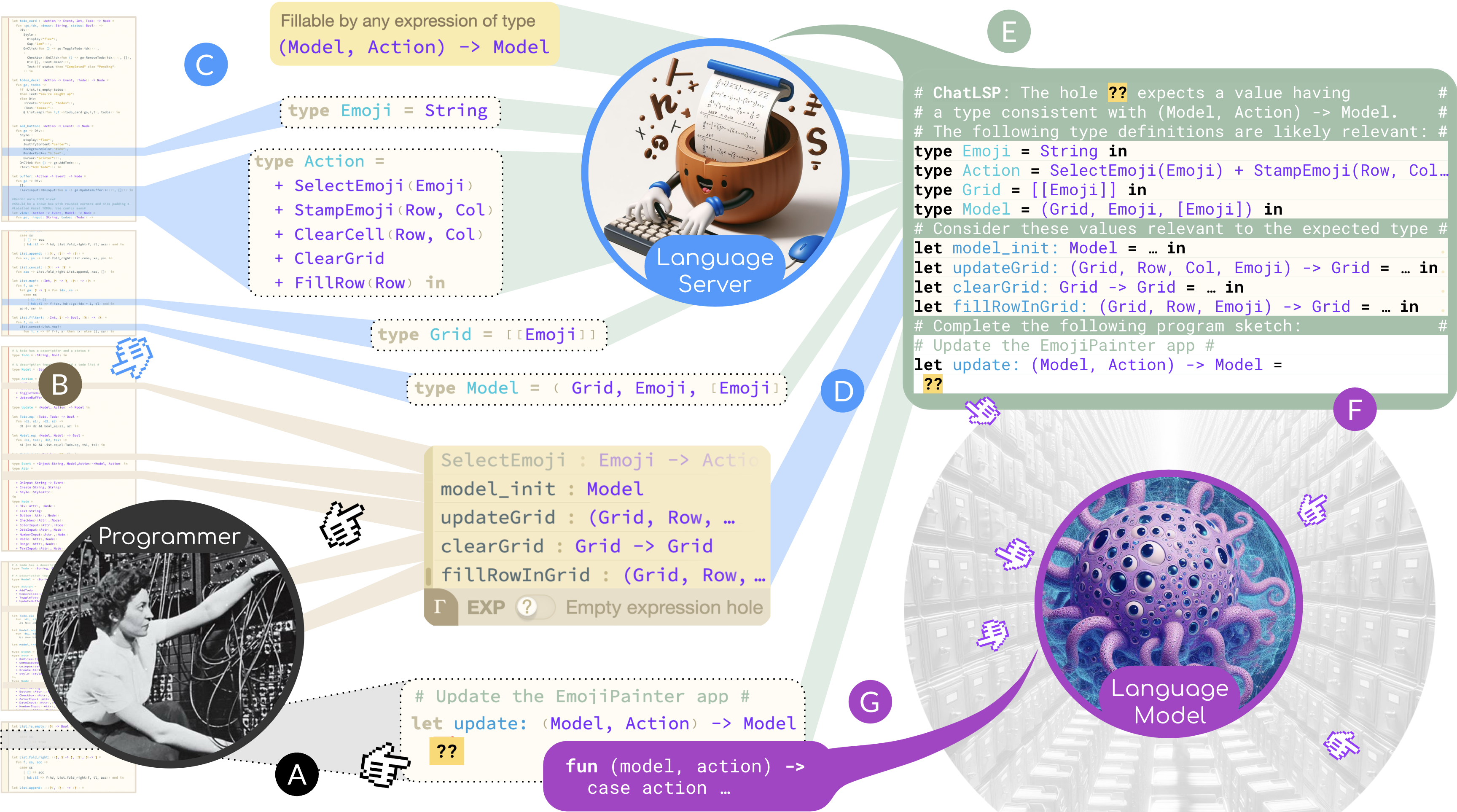}
  \caption{A programmer requests a hole filling (A) by typing \highlightli{??}, either intentionally or in a fit of frustration. The Hazel Language Server provides codebase-wide (B) semantic information relevant to the hole, collecting types based on the expected type (C) and selecting type-relevant headers from the context (D). These are combined into a contextualized text prompt (E) which is sent (F) to the LLM resulting in hole filling (G).}
  \label{fig:prompt-construction-1}
\end{figure}

The base EmojiPaint update function program sketch is as follows:

\begin{lstlisting}[style=ocamlboxed]
(* Update the EmojiPaint app model based on an action *)
let update: (Model, Action) -> Model = ?? in
\end{lstlisting}

\vspace{12pt}
We augment this sketch with additional static information obtained via the Hazel Language Server, serialized into text, displayed as a kind of projected view of the codebase -- a static program slice contextualized to the relevant program hole. Specifically, we retrieve the following static information (diagrammed in \autoref{fig:prompt-construction-1}):
\begin{itemize}
    \item \emph{Type Retrieval}: The expected type at the cursor, along with the definitions of any types aliases occurring in that type, and the definitions of aliases occurring in that definition, and so on recursively until we arrive at base types.
    \item \emph{Header Retrieval}: A selection of values, annotated with their types, filtered from the typing context based on a type-directed metric of relatedness to the expected type described below.
\end{itemize}

While we use Hazel to illustrate our approach, our goal is to outline an approximate API which could be implemented by any language server for a typed language which could drive a similar system in another language. We'll define the approximate methods for such an API as we go, and later collect them in \autoref{sec:chatlsp}.

\subsubsection{Relevant Type Definitions}

Given the above program sketch, the expected type of the hole \highlightli{??} is \codetype{(Model, Action) -> Model}. While in this example, adding the expected type to the prompt is strictly redundant, as it already appears as the function's type annotation, in general Hazel's bidirectional type system~\cite{HazelnutPOPL,hazel-popl24} allows a similar expected type to be extracted in any position for which there exists type constraints, such as in function argument position, or in a module signature including \li{update}.

Absent this sort of context, this type is elucidatory on its own. Based on the provided comment, a language model might and likely will `guess' that these refer to the state and state changes of an Model-View-Update application. But as we shall see, it is unlikely to guess the precise structures of the types the programmer has actually used. An example demonstrating the common case is show in \autoref{fig:emojipaint-bad-update-1}:

\begin{figure}[h]
  \includegraphics[width=10.7cm]{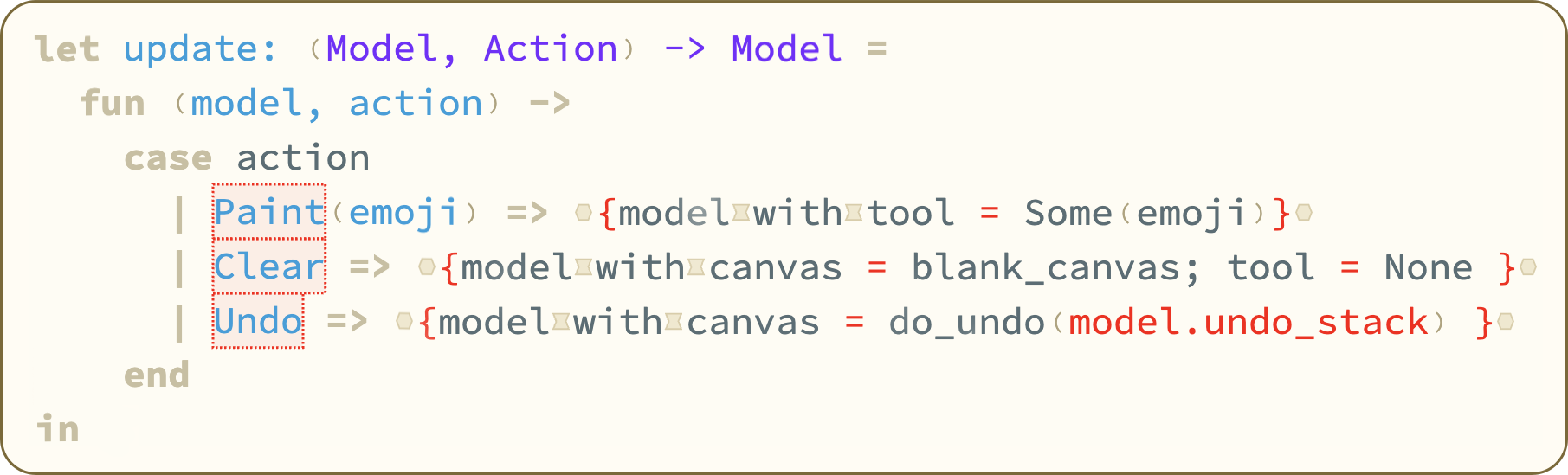}
  \caption{A typical completion with no static retrieval. Here, the language model hallucinates plausible but incorrect constructors for the Action type, and hallucinates Model as a record type using syntax not supported in Hazel.}
  \label{fig:emojipaint-bad-update-1}
\end{figure}
Hence we do automatically what a programmer in an unfamiliar codebase might do manually: recursively pursue type definitions to unwind the local semantic context hinted at by the type expectation. Providing this list to a language model is analogous to a human using the IDE to hover over types and jump iteratively to their definitions. First, we extract relevant type aliases:

\begin{itemize}
    \item \codetype{Model}
    \item \codetype{Action}
\end{itemize}

Then, we retrieve their definitions:

\begin{itemize}
    \item \li{type} \codepat{Model} \li{ =} \codetype{(Grid, Emoji, [Emoji])}
    \item \li{type} \codepat{Action} \li{ =} \codetype{SelectEmoji(Emoji) + StampEmoji(Row, Col) + ...}
\end{itemize}

And finally, we transitively complete this process, retrieving any additional aliases which occur in those definitions: the aliases \codetype{Grid} and \codetype{Emoji} from the \codetype{Model} definition and the aliases \codetype{Row}, and \codetype{Col} from the \codetype{Action} definition:

\begin{itemize}
    \item \li{type} \codepat{Emoji} \li{ =} \codetype{String}
    \item \li{type} \codepat{Row} \li{ =} \codetype{Int}
    \item \li{type} \codepat{Col} \li{ =} \codetype{Int}
    \item \li{type} \codepat{Grid} \li{ =} \codetype{[[Emoji]]}
\end{itemize}

Note that even though these type definitions are, from the point of view of the Language Server, abstract entries in the typing context, we are specifically electing to reproduce them in format resembling their original lexical concrete syntax. We have noticed that attempting this kind of naturalistic reproduction increases that chance that language model generations stay on-task, generating code in the relevant concrete syntax without reverting to prose or ill-formed syntax. 

To support the above method, a language server could implement the following methods, which may be implementable as thin glosses on top of existing methods such as 'Go to type definition':

\vspace{5pt}
\begin{itemize}
    \item \li{getExpectedType:} \codetype{(Program, LexicalLocation) -> Type}
    \item \li{extractAliases:} \codetype{Type -> [Type]}
    \item \li{getTypeDefinition:} \codetype{TypeAlias -> Type}
\end{itemize}
\vspace{5pt}

\noindent
To briefly contrast this process to embedding-vector-based retrieval augmented generation:
\begin{itemize}
    \item While vector retrieval might flag these definitions as \emph{possibly} relevant, note that they are \emph{necessarily} relevant. Our knowledge of the language semantics means we know that any completion \emph{must} respect these types, an assurance which allows us to offload burden from the more expensive and imprecise associative lookup.
    \item This necessary relevance also increases the chance that subsequent recursive retrievals will be relevant, addressing the issue of reliable \emph{multi-hop lookups} noted by  industry implementers \cite{cursorProblems2024} 
    \item Static retrieval necessarily respects scope. Vector retrieval may return related-seeming definitions, but without exact knowledge of the language semantics, there is no guarantee that these will be \emph{the} definitions relevant to this lexical context.
\end{itemize}

\subsection{Relevant Headers from the Typing Context}

In addition to relevant type definitions, we augment the prompt with the names and types (which together we term the \emph{headers}) of relevant values -- typically functions -- from the typing context. From a user interface perspective, this is analogous to a type-directed autocomplete menu.

Our extraction method divides into three stages:
\begin{enumerate}
    \item Use the expected type to identify a list of target types
    \item Filter the typing context for values with types related in a certain way to these target types
    \item Assign scores to each element of the resulting list, and return the prefix of that list truncated at some scoring and length thresholds (here, score > 0.0 and 10 items respectively)
\end{enumerate}

The resulting context entries are formatted as code sketches, again to facilitate language model ingestion. For example the pair (\li{string_of_int},  ~\codetype{Int -> String}) is formatted as:

\vspace{6pt}
\begin{lstlisting}[style=ocamlboxed]
    let string_of_int : Int -> String =  in   
\end{lstlisting}
\vspace{6pt}

Here, the body of the definition is simply omitted. Interestingly, we originally used ellipsis ($...$) in place of the body, but this resulted in an increased chance the model (especially the smaller StarCoder2 model) would itself emit the token $...$ in lieu of a full completion.

\subsubsection{Identification of Target Types} \label{identification-of-target-types}

First, we deconstructing the expected type to identify relevant sub-components which could be used, in conjunction with their relevant elimination forms, to construct the target. Our initial target type is simply the type of the hole itself: 

\vspace{6pt}
\textbf{Target types} = \codetype{(Model, Action) -> Model}
\vspace{6pt}

Then, if that type is a compound type, we consider its components. In particular, if the type is a product type, we consider its components to be targets, and if the type is an arrow type (as it is here), we consider its return type to be a target.

\vspace{6pt}
\textbf{Target types} = \codetype{ (Model, Action) -> Model, Model, ...}
\vspace{6pt}

In principle, we could continue this deconstruction recursively indefinitely, but for our immediate purpose of identifying likely-relevant types, we've found it suffices to extend one more iteration; that is, product/arrow types containing product/arrow types. 
It simplifies our calculations to internally normalize all type definitions (Hazel is structurally typed). Here, we will only do so opportunistically for clarity of presentation:

\vspace{6pt}
\textbf{Type of hole} = \codetype{ ((String, (Grid, Emoji, [Emoji])), Action) -> (Grid, Emoji, [Emoji])}

\textbf{Target types} = \codetype{ (Model, Action) -> Model, Model, Grid, Emoji, [Emoji]}
\vspace{6pt}

One can likely see other ways of extending target type extraction. Possibilities included destructuring more compound types such as records, or for function types, also considering the input types as a kind of negative target, in that we may want to prioritize types that consume a relevant type from the local context. For now we proceed with the simple approach outlined.

We do not, however, return unaliased base types such as \codetype{Bool} or \codetype{String} as target types. Early experimentation indicated that, given that there are typically many standard library functions on base types, often with no a priori way to distinguish their relevance based on types, such functions would often act as confounders, since which happened to be included was incidental. In practice, a standard library would already be well-understood by an LLM from the pre-training or fine-tuning step. For Hazel, we replicate much of the OCaml standard library to sidestep this need.

\subsubsection{Filtering the Context}

For each target type, we filter the typing context to retrieve types which can be used, again in conjunction with appropriate elimination forms, to produce the target.

This is essentially similar to target type extraction. In particular, we return types which are consistent with the target, arrow types whose return type is consistent with the target, and product types whose have a component consistent with the target. For example:

\vspace{6pt}
\textbf{Target type} \codetype{Grid} \textbf{yields}
\begin{itemize}
    \item \li{updateGrid:} \codetype{(Grid, Row, Col, Emoji) -> Grid}
    \item \li{clearGrid:} \codetype{Grid -> Grid}
    \item \li{fillRowInGrid:} \codetype{(Grid, Row, Emoji) -> Grid}
\end{itemize}
\vspace{6pt}

\subsubsection{Sorting and scoring the filtered context}

Prior work \cite{vlhcc2022assistant} has surveyed various ways in which semantic information can be used to sort typing-context-originating suggestions for relevance. For our purposes here we use a simple scheme intended as a proof-of-concept to establish a baseline for more sophisticated methods.

By default, Hazel context entries are sorted by locality of definition, which provides a reasonable default for relevance. Thus we sort stably with respect to the locality ordering for context entries having the same score.

By default, all entries are assigned a score of 1.0. However, if a type contains the (gradual) unknown type (\codetype{?}), a multiplier is applied based on the ratio of unknown to known type constructors in the type (for example, the type \codetype{[?]} -- a list of unknown type -- would receive a multiplier of 0.5, since the list constructor is known). This acts simply to de-prioritize incomplete implementations, about which not enough information is available to make it a good suggestion.

Here are the relevant headers from the EmojiPainter example:

\vspace{6pt}
\begin{lstlisting}[style=ocamlboxed]
"Consider using these variables relevant to the expected type:" 
let model_init: Model =  in
let fillRowInGrid: ((Grid, Row, Emoji) -> Grid) =  in
let clearGrid: (Grid -> Grid) =  in
let updateGrid: ((Grid, Row, Col, Emoji) -> Grid) =  in
\end{lstlisting}
\vspace{6pt}

In order to implement relevant header extraction in an arbitrarily language server, one could provide the following methods:

\vspace{6pt}
\begin{itemize}
    \item \li{getTargetTypes:} \codetype{Type -> [Type]}
    \item \li{filterContext:} \codetype{Context, Type -> [(Name, Type)]}
    \item \li{scoreEntry:} \codetype{(Name, Type) -> Float}
\end{itemize}

\subsection{Syntactic and Semantic Error Correction}
\label{sec:error-correction}
The use of instruction-tuned language models makes available a lightweight form of program repair based on an iterative loop of generating completions and retrieving error messages from compilers or static analyzers. The general technique of looping LLM code generation on compiler errors appears to have emerged in tandem with early LLM code generation experiments \cite{VulnerabilityRepair} \cite{zhang2022repairing} and has been examined in greater detail by \citet{joshi2022repair}.

After receiving a response from the model, we substitute the received completion into the original program sketch. We then query the Hazel Language Server to parse the resulting program. Hazel parsing is strongly incremental, enabling the (partial) type-checking of programs even in the presence of unrecognized or missing delimiters. We then query the language server for a list of static errors, which include syntax and type errors. If there are any such errors, we serialize them to a string, and send them to the language model.

In order to maintain model context, we append the errors to a growing log of messages beginning with the original prompt. The number of correction rounds which can be performed in this way is thus limited by the length of the context window; in our case, using the 8k token window of GPT4-0613, we are effectively capped at 5 rounds. However, we have noticed that 2 rounds are often sufficient to eliminate static errors, and that rounds in excess of 2 tend to show diminishing returns, so we have capped the maximum number of rounds at 2.

To support this in another language, its language server must be able to localize static errors, reporting locations and error messages. Ideally, it would produce a list of errors, rather than just the first error encountered, as is supported by Hazel's total type error localization and recovery system~\cite{hazel-popl24}. This could be achieved by implementing the following method:

\vspace{6pt}
\begin{itemize}
    \item \li{getStaticErrors:} \codetype{Program -> [StaticError]}
\end{itemize}

\subsection{Experimental Evaluation}

We now evaluate the effectiveness of this method of proactive static contextualization and retrospective correction for LLM code completion.

\subsubsection{MVUBench}
\label{sec:mvubench}
Hazel is a low-resource language, so we are unable to conduct an at-scale evaluation in this context. Instead, as previously motivated, we construct a benchmark suite of five MVU applications including the EmojiPainter example from the previous sections:
\begin{itemize}
    \item \textbf{Todo (\emph{TO})}: Maintains a list of tasks
    \item \textbf{Room Booking (\emph{BO})}: Manages a room booking schedule
    \item \textbf{Emoji Painter (\emph{EM})}: Paints emoji stamps on a small canvas
    \item \textbf{Playlist Manager (\emph{PL})}: Manages a music playlist
    \item \textbf{Password Strength Checker (\emph{PA})}: Rates a password via a dynamic set of properties
\end{itemize}

The baseline program sketch provided to the language model for each of these programs is simply the type-annotated function header for its corresponding \li{update} function, along with a single-line comment describing that function's purpose, including the name of the application, in line with the running example.

Each application also comes with a simulated repository containing relevant (and less relevant) type and utility function definitions. 

We also provide a small test suite for each example, consisting of 10-15 tests ensuring that each MVU action behaves as a user might reasonably expect without additional specification.

In such a situation, a naive language model completion would be informed only by the \li{update} function type aliases (which are often generic terms such as \li{Model}) and the single-line comment (which only hints at the intended functionality). While it is still possible that in very typical situations, the model might correctly guess appropriate types and names, more likely (as we shall see) it will hallucinate plausible-but-incorrect completions. By varying the methods through which additional context is provided, and the corrective methods applied to resulting completions, we provide a baseline analysis for the relative effects and interactions of these methods on LLM code completion.

\subsubsection{Feature Ablation Experiment}

Our main experiment consisted of 320 completions trials, each of which makes between one and three calls to the language model. These 320 trials divide as follows:

\begin{itemize}
    \item \textbf{8 feature ablation configurations}
    \begin{itemize}
        \item \emph{Type Retrieval}: Whether to include expected type and type definitions
        \item \emph{Header Retrieval}: Whether to include relevant headers from the typing context
        \item \emph{Error Rounds}: Whether to perform up to 2 static error correction
    \end{itemize}
    \item \textbf{5 program sketches (TO, BO, EM, PL, PA)}
    \item \textbf{20 completion trials per combination} (to account for model non-determinism~\cite{ouyang2023llm}). We ran these experiments at temperature 0.6 (a hyperparameter effecting the stochasticity of token sampling), selected based on trial experiments as a balance between noisy variance and producing a range of interestingly distinct completions
\end{itemize}

\subsubsection{Comparison Baseline 1: No Context}
The ablation configurations lacking all static retrieval feature serves as a lower bound baseline -- without any context except the brief comment on the \li{update} function, we would expect even high-performing models to perform poorly due to lack of context. This is the current reality for AI programming assistants that do not attempt repository-level retrieval.

\subsubsection{Comparison Baseline 2: Exhaustive Retrieval}
An additional baseline configuration, beyond those outlined above, is to perform exhaustive retrieval of all application code, excluding tests, up to the context window limit. This serves as a token-inefficient upper bound on performance.

\subsubsection{Comparison Baseline 3: Vector Retrieval with Confounds}
\label{sec:vector-RAG}
Finally, we compare our approach to vector retrieval. Given that our test corpus consists of 5 relatively small programs, we have emulated a larger more realistic codebase by combining those five programs, minus tests and update functions, along with Hazel's standard library, to create a 1000-line simulated codebase from which context can be drawn. Combining these programs has the effect of creating some possible lexical confounders, e.g. two types having the same name; we contend that this construction, albeit synthetic, nonetheless emulates a legitimate source of confusion for a scope-unaware method like vector retrieval. 

We have used the simplest standard RAG strategy, uniformly dividing the codebase into 150-character chunks, which were submitted to OpenAI's Ada (text-embedding-ada-002), a commercial embeddings model \cite{embeddings}. The retrieved 1536-element vectors, along with their associated text chunks, are then stored locally in a JSON file acting as a basic vector database.

In order to retrieve chunks relevant to a provided sketch (our function headers and comments), we submit that sketch to the same API endpoint, and then search our vector database for the top 6 chunks with the highest cosine similarity \cite{embeddings}.

The above parameters (150 character chunks, 6 entries) are chosen so that the total (900 characters) lines up with the average length of the total static retrieval context (types + relevant context) for our 5 examples, with the chunk size being set as small as possible while still being able to fully contain most type definitions in our corpus.

It should be noted that there exist a variety of more advanced chunking strategies which may yield better results, including overlapping windows, chunks aligned to inferred authorial intent \cite{wang2024rlcoder}, and semantic chunking which takes into account source syntax. However, all these strategies have complex trade-offs which take us beyond our immediate comparative goals; for example, chunking by top-level definitions (a language-aware approach) might prevent issues with a poorly-truncated definition being included in a prompt, but seeing as definitions can range widely in size, being forced to include an entire definition may prevent multiple chunks which are together more relevant from being included.

As such, we have elected to leave the RAG baseline structurally agnostic, so as to more cleanly contrast it with semantic methods, while noting it is likely that ultimately these two methods are not exclusive and can be used synergistically in a production setting (for example, balancing the ratio of typed semantic versus associative RAG depending on the amount of static information available at a given lexical location). We return to this theme in \autoref{sec:related-work}.

\subsection{Hazel GPT-4 Results}
\begin{figure}[h]
  \includegraphics[width=13cm]{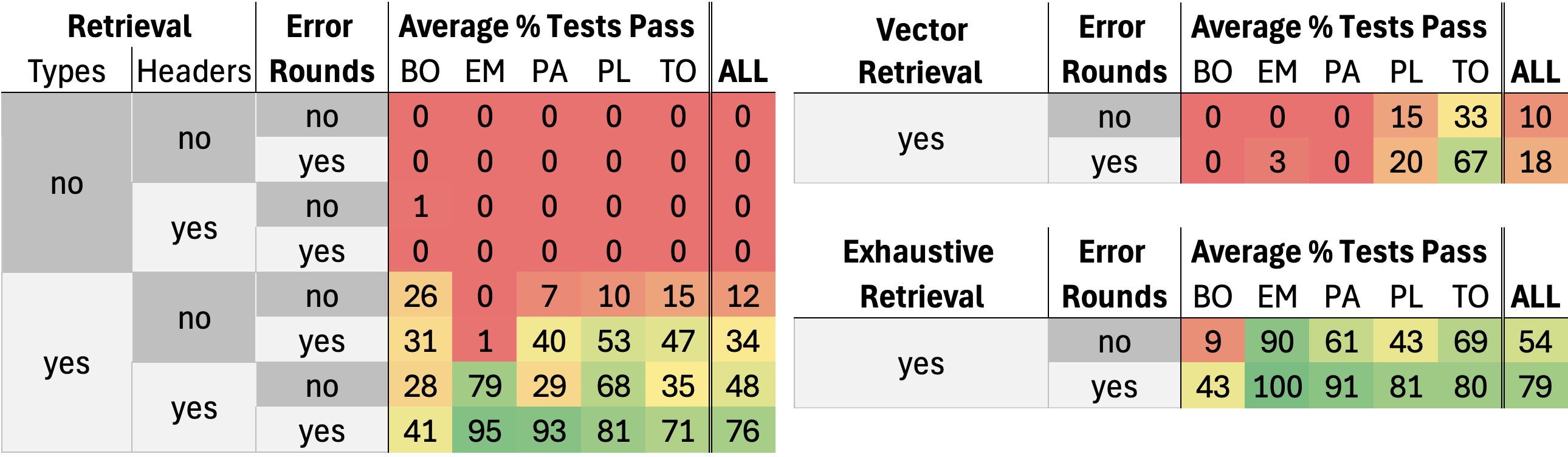}
  \caption{Hazel GPT-4: Results for guided completion (20 trials per, temperature 0.6)}
  \label{fig:hazel-gpt4-bigexp-0.5}
\end{figure}

\noindent
\autoref{fig:hazel-gpt4-bigexp-0.5} shows the results of our evaluation for GPT-4. We see a clear trend of more semantic information yielding better results on the held-out tests. The no-context baseline (no/no/no) alone does not suffice to yield meaningful generations, often returning syntactically incorrect code as the model hallucinates data types and syntax (such as OCaml-style records) which do not exist in Hazel, despite the inclusion of the Hazel Crash Course.

Including type definitions seem absolutely necessary to allow the model to scaffold the update function. Without this scaffolding, relevant function headers alone show little effect on correctness. However, in combination with types, function headers have a large multiplicative effect, increasing test performance threefold. \autoref{fig:emojipaint-sample-results-1} provides concrete examples of this interaction:
\begin{figure}[h]
  \includegraphics[width=\linewidth]{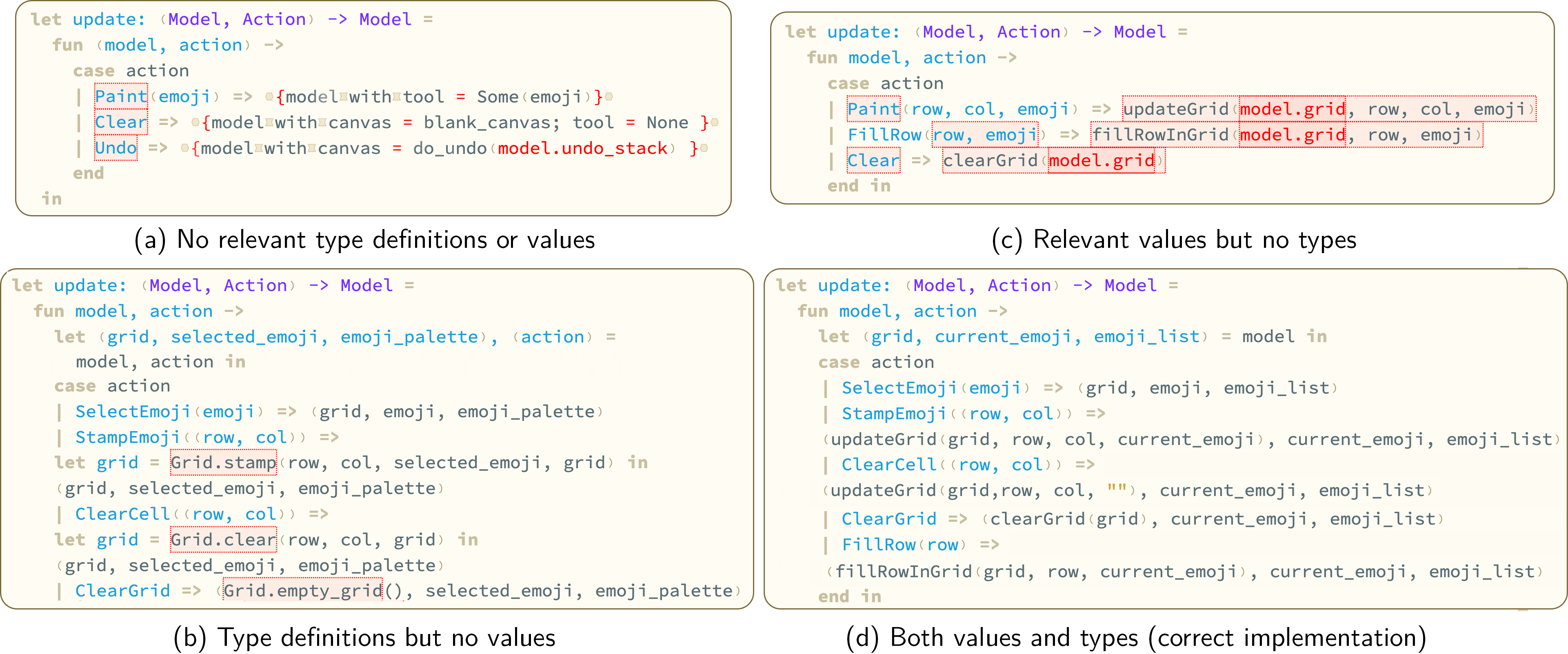}
  \caption{Some sample completions for various configurations. In (a), without any supporting context, we see reasonable-but-incorrect guesses at both the \li{Action} constructors and the \li{Model} type; furthermore, the model completion uses record syntax which does not exist in Hazel. (b) uses the provided types correctly but hallucinates helper names. In (c) we see uses of appropriate helpers, but (mostly) incorrect guesses for \li{Action} constructors. (d) exploits the provided context to produce a fully correct solution.}
  \label{fig:emojipaint-sample-results-1}
\end{figure}

\begin{figure}[h]
  \includegraphics[width=\linewidth]{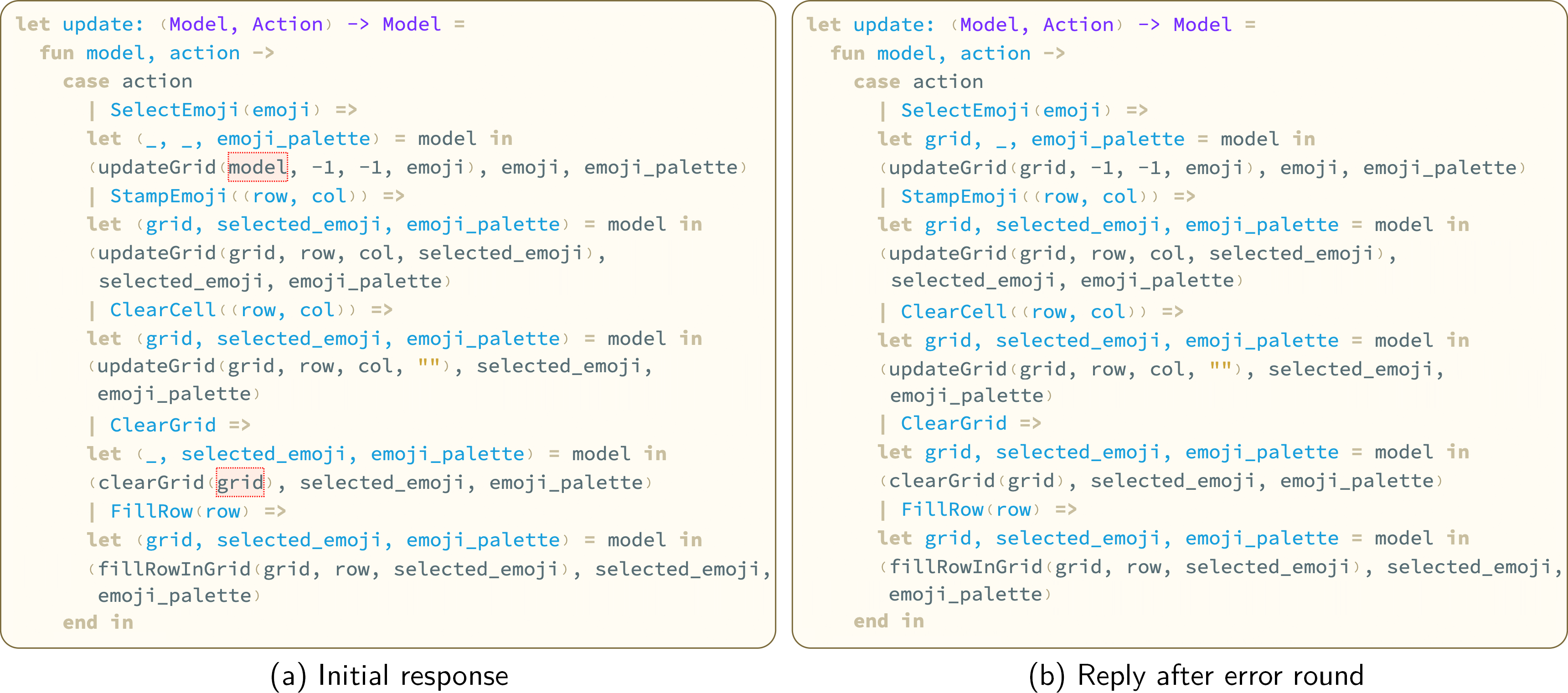}
  \caption{Error rounds were generally very effective at correcting an almost-correct program. Here, the error round reply included a type inconsistency error on \li{model} and an unbound variable error in \li{grid}, both of which were corrected in the model's reply}
  \label{fig:emojipaint-error-rounds-1}
\end{figure}

Similarly, error rounds on their own are ineffective on code consisting largely of hallucinated types and functions. But given the scaffolding effect of relevant static information, they act multiplicatively, increasing performance by a factor of 4 for types without headers, and a factor of 1.5 for both types and headers. Error rounds were particularly effective at transforming almost-correct completions to fully correct ones, as shown in \autoref{fig:emojipaint-error-rounds-1}.

\begin{figure}[h]
  \includegraphics[width=\linewidth]{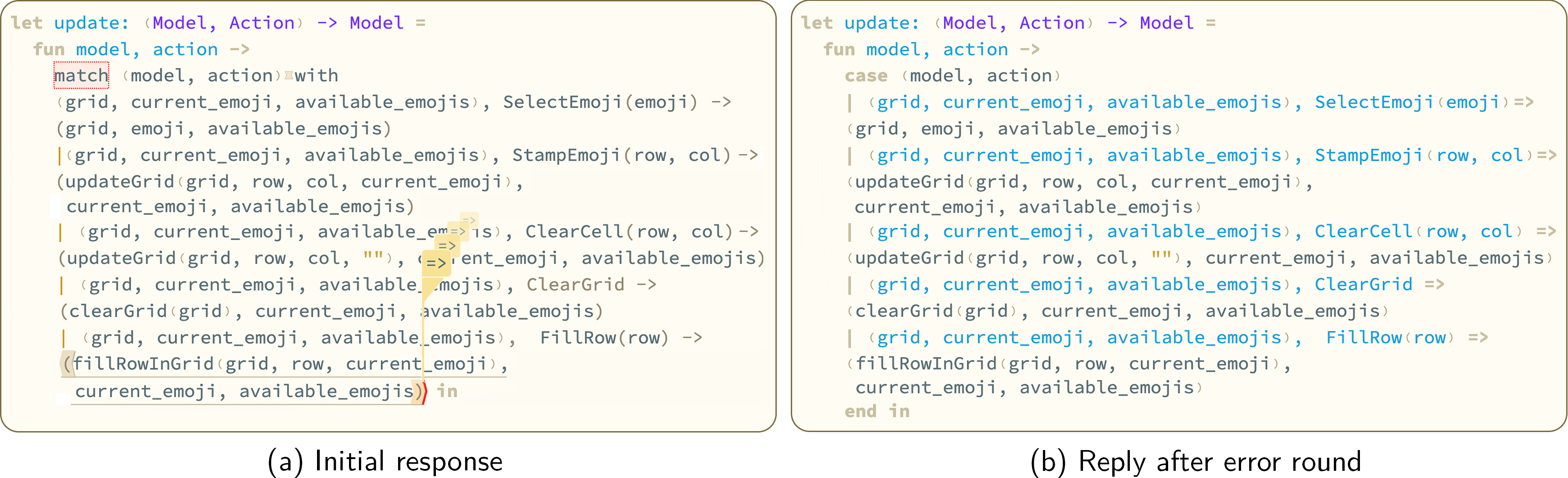}
  \caption{Here, a parse error (match/with used instead of \li{case}) was corrected, even though the Hazel error message in this case is somewhat unclear: \emph{``The parser has detected unmatched delimiters: \highlightli{=>}, \highlightli{=>}, \highlightli{=>}, \highlightli{=>}. The presence of a \highlightli{=>} in the list likely indicates that a \highlightli{->} was mistakenly used in a case expression.''}}
  \label{fig:emojipaint-error-rounds-2}
\end{figure}

One phenomenon of note was that sometimes even poor error messages proved effective; the error in \autoref{fig:emojipaint-error-rounds-2} is an at-most partially accurate characterization of the syntax error, but knowing there was a syntax error proved sufficient for the model to correct it, perhaps due to the additional context provided in the Hazel Crash Course.

\begin{figure}[h]
  \includegraphics[width=5cm]{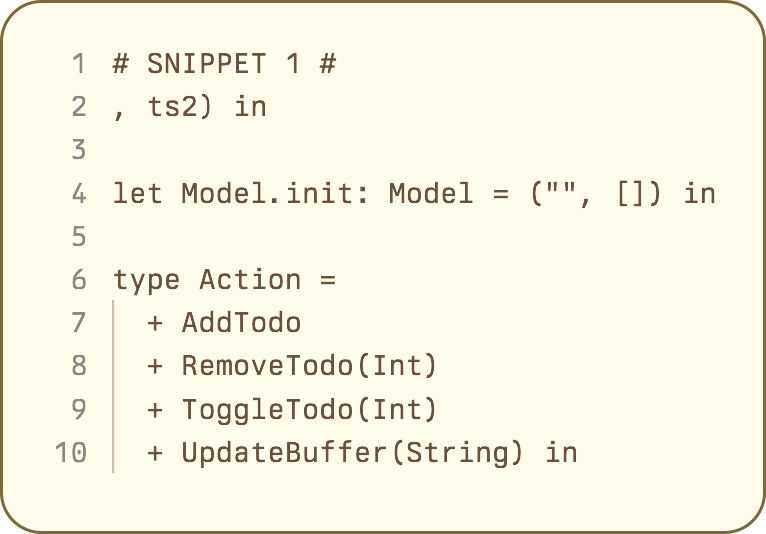}
  \caption{A confounding snippet commonly retrieved by vector retrieval}
  \label{fig:RAG-confounder}
\end{figure}

The combination of types and headers performed well against the Vector Retrieval baseline, though it should be noted that this was disproportionately due to a single confounding chunk which was retrieved for each example, even though it is only relevant to the Todo application (see \autoref{fig:RAG-confounder}). We believe that, due to the fact that this chunk coincidentally includes the word symbol \codetype{Model} twice alongside \codetype{Action}, it is deemed relevant to each \li{update} sketch. The inclusion of this snippet often resulted in the language model attempting to implement a Todo application, or some hybrid thereof. We debated refining the chunking strategy to remove this confounder, but found it easy to inadvertently create similar scenarios; ultimately, it is a representative artefact of a process which is fundamentally non-scope-aware. See \autoref{section:threats-to-validity} for further discussion.

\begin{figure}[h!]
  \includegraphics[width=\linewidth]{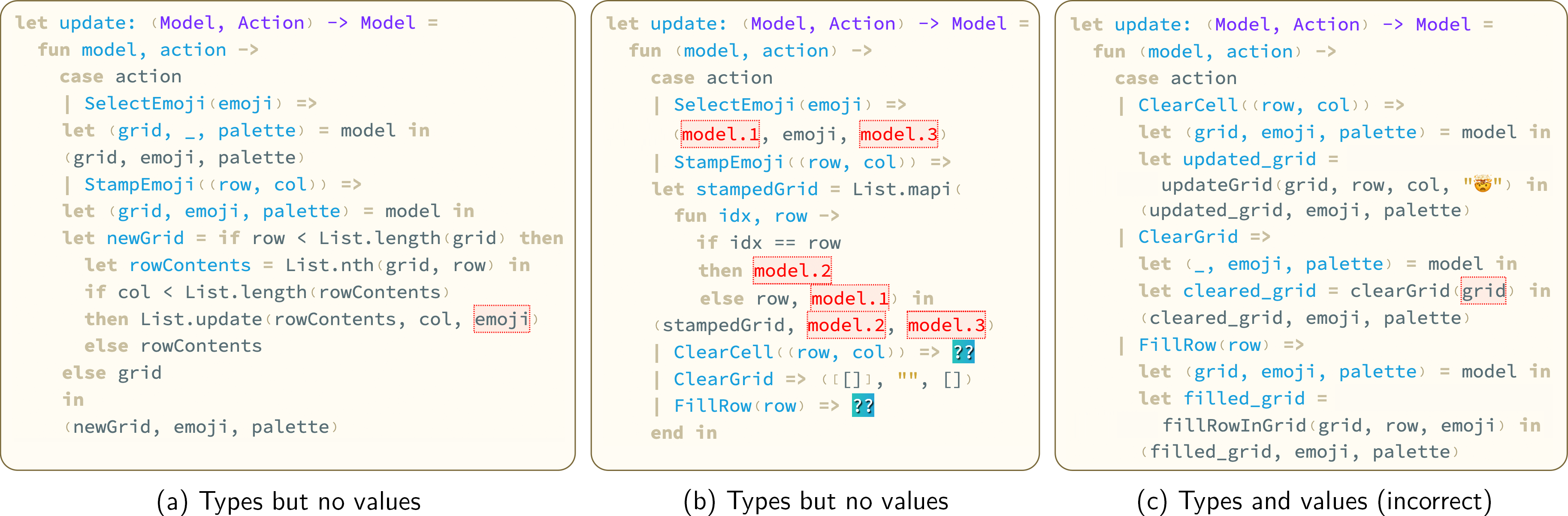}
  \caption{Some more exotic completions: (a) is an excerpt of a very long solution which, not having access to headers, almost successfully managed to re-implement the required logic in-line. In (b) we see an example of a completion containing explicit holes, likely due to their presence in our few-shot example sketches. In (c), the LLM somewhat quixotically suggests substituting a different emoji in lieu of the empty string.}
  \label{fig:emojipaint-sample-results-2}
\end{figure}

See \autoref{fig:emojipaint-sample-results-2} for additional examples of more atypical completions. 

It is worth nothing that although the results for types + headers are similar to those for exhaustive retrieval, with exhaustive retrieval performing somewhat better, our experiments are not powerful enough to significantly distinguish between these cases, as the sizes of our programs are small enough that the context size delta is not reflective of real-world use cases. Specifically, the size of the retrieval context averaged 890 characters for our programs, whereas the exhaustive context averaged 1370 characters. As it stands the performance delta seems to compare positively to the context length (and hence cost) delta, but more data is needed to make this conclusive.

\subsubsection{Token and Time Performance}

\begin{figure}[h]
  \includegraphics[width=12.5cm]{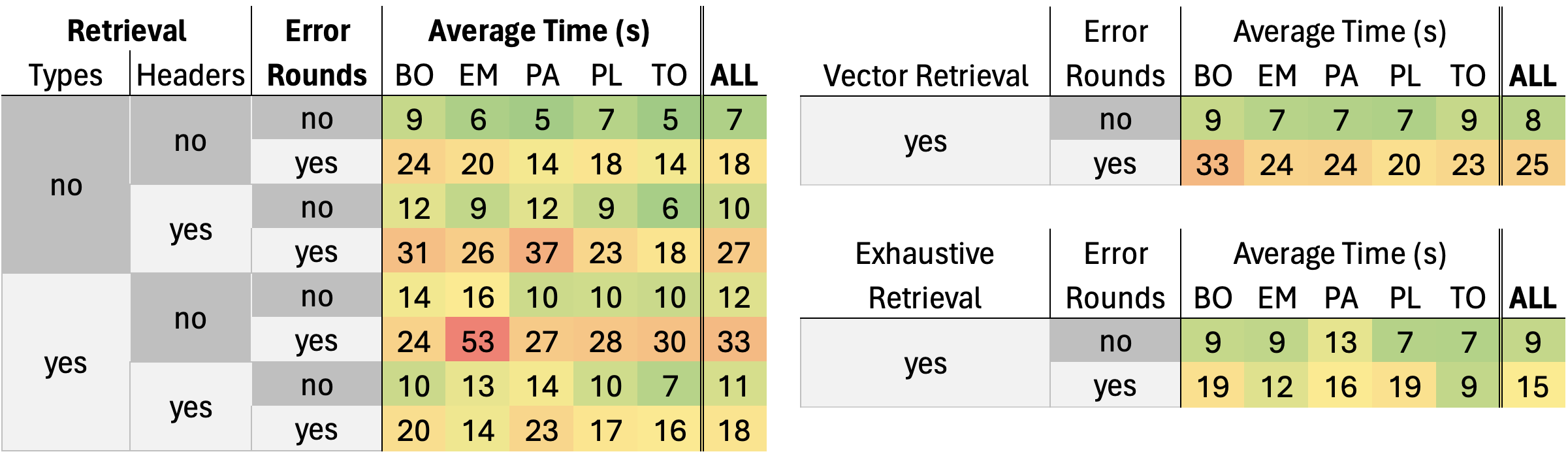}
  \caption{Hazel GPT-4: Time elapsed for guided completion (20 trials per, temperature 0.6)}
  \label{fig:hazel-gpt4-time-taken}
\end{figure}

\autoref{fig:hazel-gpt4-time-taken} shows the time taken in seconds for all trials. The time taken is dominated by the number of round trips through the API, with each round scaling in proportion to the sum of the length of the context and the length of the generation. Generally these times are too long for use in a practical completion setting; our intention is to determine a ceiling on current performance with respect to correctness rather than present a practical system. Note however that these times will likely decrease quickly with hardware and software advances. As of May 2024 GPT-4o\cite{gpt4o} performs on average twice as fast as the GPT-4-0613 model checkpoint used for our experiments. However, the long worst-case times for error rounds suggests that capping at a single correction round may be more practical, or motivate the use of summarization to reduce token count during error rounds.

\begin{figure}[h]
  \includegraphics[width=\linewidth]{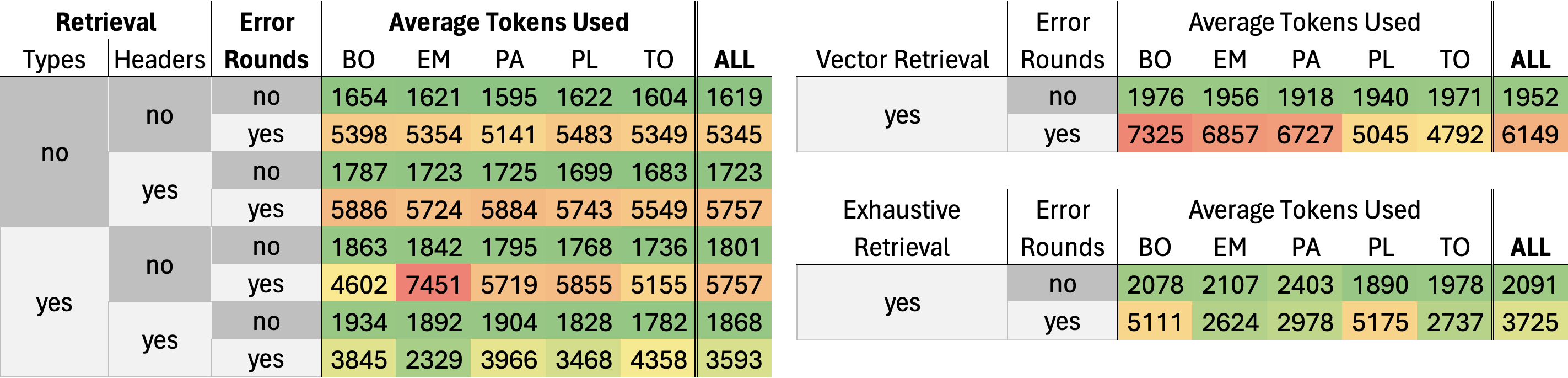}
  \caption{Hazel GPT-4: Tokens used for guided completion (20 trials per, temperature 0.6)}
  \label{fig:hazel-gpt4-tokens-used}
\end{figure}

Figure \autoref{fig:hazel-gpt4-tokens-used} shows the total tokens used, both sent and received from the API. These are roughly proportional to the time taken, and precisely proportional to the total cost.

\subsection{Hazel StarCoder2-15B Results}

\begin{figure}[h!]
  \includegraphics[width=10.6cm]{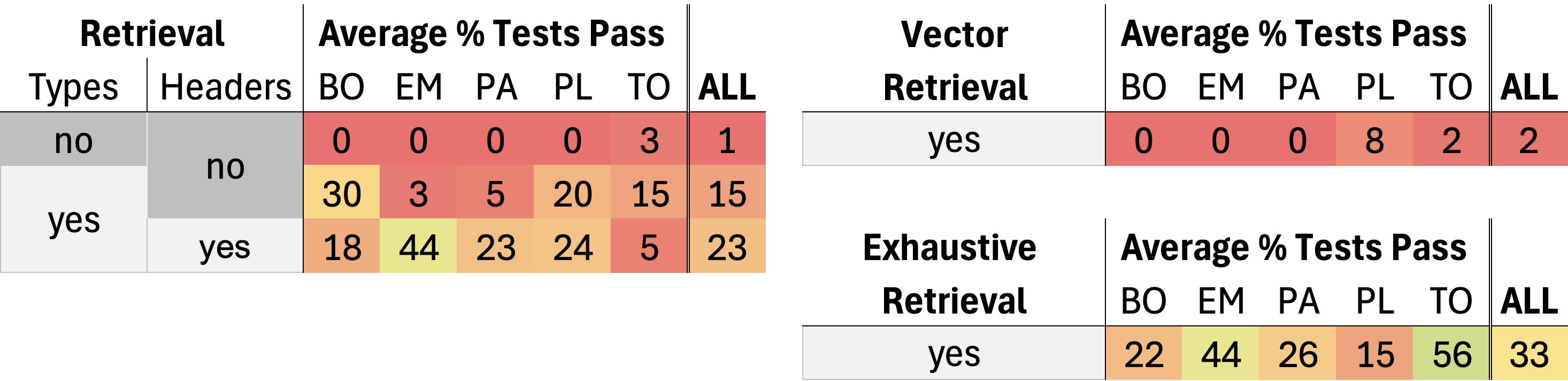}
  \caption{Hazel StarCoder2: Results for guided completion (20 trials per, temperature 0.6)}
  \label{fig:hazel_starcoder2}
\end{figure}

To assess the effectiveness of static retrieval with smaller completion models, we conducted tests using StarCoder2-15B, a model small enough to be run locally on consumer hardware. The average percentage of tests passed, shown in the rightmost column of \autoref{fig:hazel_starcoder2}, exhibits a consistent trend with the GPT-4 results. In the absence of any type or header information, StarCoder2 performed poorly. The addition of type information drastically improves performance, increasing the percentage of correct solutions by an order of magnitude. Furthermore, incorporating headers leads to an additional 50\% increase in relative performance.

However, two examples, \textbf{BO} and \textbf{TO}, experienced degraded performance after the inclusion of headers. After close examination of the headers and the output programs, we discovered that the completions tending to use type-appropriate but in-fact irrelevant retrieved headers. We hypothesize that smaller completion models, such as StarCoder2, are more sensitive to code that appears near the end of the context window, making them more susceptible to the influence of irrelevant information. We touch on this failure mode again when we consider related work in \autoref{sec:related-work}.

Vector retrieval baseline performance was significantly worse (in absolute and relative terms) than with the larger model. We conjecture that this is due to a heightened sensitivity to erroneous syntax in the prompt created by chunk truncation. 

\section{Static Retrieval in TypeScript}
\label{sec:retrieval-typescript}

To confirm that the above results are not an artefact of using a low-resource language, we also experimented with static retrieval in TypeScript. 

\subsection{TypeScript Methodology}

\begin{figure}[h]
  \includegraphics[width=10cm]{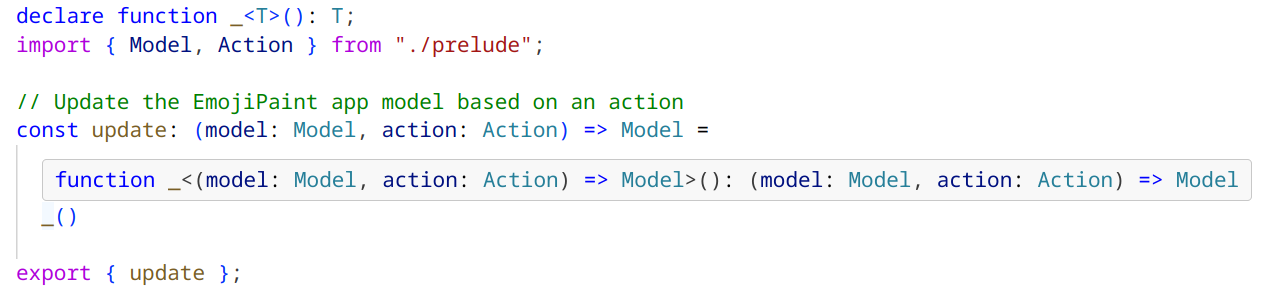}
  \caption{Hovering over a simulated program hole in TypeScript}
  \label{fig:ts-lsp-hole-hover}
\end{figure}

Our methodology roughly follows the Hazel experiments. As TypeScript is a high resource language, well-represented in training sets, we did not need to provide a syntax crash course as we did for Hazel. TypeScript lacks the explicit support for typed holes and thus a convenient way to extract semantic information. We emulated typed holes using a previously established approach~\cite{Canti_2019} using generic functions (\autoref {fig:ts-lsp-hole-hover}).

Specifically, we prefix the sketch file with the declaration: \li{declare function _<T>(): T}. Then, we represent a program hole as an application of that generic function: \highlightli{_()}. Calling the TypeScript language server's hover method on the hole gives us a corresponding type signature. It should be noted that this method of emulating typed holes is not fully general. While it works consistently for holes replacing the bodies of function definitions, it fails in some syntactic positions, including as an operand of infix operators.

Static retrieval of type definitions is performed via the TypeScript language server. In particular we use coordinated calls to the \emph{Go to Type Definition} and \emph{Hover} methods to recursively retrieve relevant types from the source lexically.

There does not appear to be any direct way of retrieving a typing context given a lexical location, or even a complete list of variables in scope using the TypeScript language server. We experimented with different methods to retrieve relevant headers, including scanning the repository using CodeQL, but were did not find a fully satisfactory general approach. Rather than incurring the engineering cost of a compiler-level intervention, we simulated the retrieval of relevant headers manually, emulating the same methodology as the Hazel Language Server. As such, our TypeScript implementation should be considered a rough proof-of-concept; our experience here motivated our prospective LSP extension outlined in \autoref{sec:chatlsp}.

We used the TypeScript compiler to collate static errors for correction rounds.

Adapting MVUBench to TypeScript was done with the aid of Claude \cite{Anthropic_2024}, an LLM chat agent (See \autoref{sec:acknowledgements} for more about our supporting LLM usage). Transliterated code was manually adjusted to establish basic conformance to TypeScript idioms, for example adding elements to array at the end, versus at the start is standard for linked lists in functional languages like Hazel. Our experience here suggests that MVUBench can be ported with relative ease to other similar languages.






\subsection{TypeScript GPT-4 Results}

\begin{figure}[h]
  \includegraphics[width=13cm]{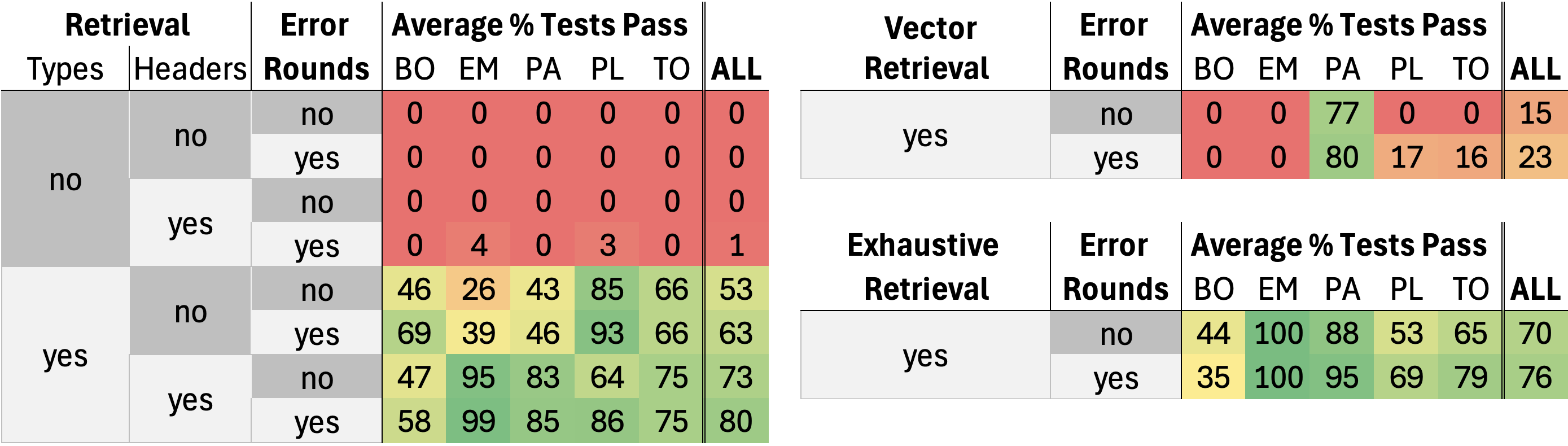}
  \caption{TypeScript GPT-4: Results for guided completion (20 trials per, temperature 0.6)}
  \label{fig:results-typescript-gpt4}
\end{figure}

In broad strokes the TypeScript results (\autoref{fig:results-typescript-gpt4}) are similar to the Hazel results. We see, somewhat unsurprisingly, that a higher-resource language, well represented in the training set, achieves better overall completions from the language model. Unlike with Hazel, some trials passed some tests even with no type information provided. With type definitions included, the TypeScript results are flatter than the Hazel results.

The ratio of tests passed with-versus-without headers is 3 for Hazel, and 1.5 for TypeScript. From examining the generated completions, we see that the model, when not provided with relevant headers, is significantly more able to produce equivalent working logic inline than it was in Hazel.

The TypeScript performance proved less dependent on error rounds. In Hazel, the ratio of tests passing with-versus-without error rounds was about 2, whereas for TypeScript it is about 1.2. Again, this is likely due to the fact that the model is far more familiar with TypeScript syntax, and unlikely to make the kind of syntax errors which the error rounds were vital for correcting in the Hazel experiment.

Performance relative to the exhaustive and vector retrieval baselines, including the high per-example variance of the latter, are relatively in line with the Hazel results.




\subsection{TypeScript StarCoder2-15B Results}

\begin{figure}[h]
  \includegraphics[width=10.6cm]{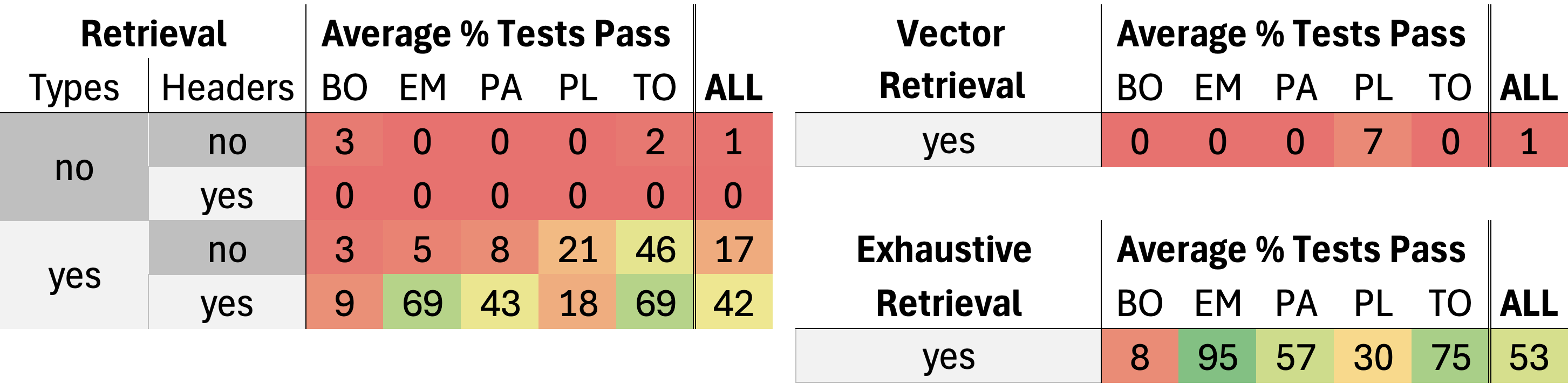}
  \caption{TypeScript StarCoder2: Results for guided completion (20 trials per, temperature 0.6)}
  \label{fig:results-typescript-starcoder}
\end{figure}

The TypeScript StarCoder2 results (\autoref{fig:results-typescript-starcoder}) appear roughly in line with the Hazel results modulo the considerations of the previous section.



\section{Threats to Validity}
\label{section:threats-to-validity}


The improvement seen from the inclusion of relevant function headers is highly contingent on the fact that many relevant functions have already been implemented. While we believe that this approximates a common case in programming practice for which naive contextualization strategies fail, validating this claim would require larger-scale study, using at-scale programs which are more neutrally selected.


More broadly, MVUBench is not (and is not meant to) be representative of all coding tasks, but rather to present a challenge to contemporary techniques and help evaluate approaches to semantic contextualization (e.g. vector retrieval, which we evaluate in \autoref{sec:vector-RAG}).

Our TypeScript MVUBench is a very close translation of the Hazel code; although the MVU paradigm is in use in the TypeScript world \cite{MVUReact}, there is a question as to the applicability of these methods to a broader range of TypeScript programming styles. 

The appropriateness of our baselines is arguable. This is an idealized setup in which the baseline case is to provide no information beyond the function header; it is unsurprising that adding related context drastically improves the result. In many real cases the cursor window would contain a large amount of relevant code. We have chosen here to focus on the situation where the window does not contain much relevant code, but it remains to validate the relative rate of occurrence of these scenarios in the wild.

Our RAG baseline is relatively simplistic, and practical implementations are increasingly integrating more sophisticated methods which may more closely approximate static retrieval. The fact that we compensated for the small size of our examples by creating a conjoined codebase to create our embedding vector database may not be adequately representative of a real large-scale codebase.

\section{ChatLSP}
\label{sec:chatlsp}

Here we sketch a conservative extension to the Language Server Protocol to support static contextualization, motivated in part the awkwardness of implementing  static contextualization in TypeScript using its existing language server. The interface differs somewhat from the API we sketched incrementally in \autoref{sec:semantic-retrieval}, as the LSP is presentation-centric, operating in terms of strings and affordances rather than language-specific semantic data types.

Immediately following, we will sketch how one might implement this \emph{ChatLSP} API in terms of our \emph{Static Contextualization} API, the latter serving more as an internal interface for language server implementers. 

\subsection{ChatLSP API Methods}

\vspace{3pt}
\begin{enumerate}
    \item \textbf{aiTutorial}: A constant (lexical-context-independent) method for low resource languages (like Hazel) to specify a textual tutorial intended for LLMs having robust support for in-context learning. For high resource languages, the default implementation will simply return a string stating which language is in use.
    \item \textbf{expectedType}: Returns a string specifying the expected type at the cursor, if available
    \item
    \textbf{retrieveRelevantTypes}: Returns a string containing type definitions that may be relevant at the cursor location
    \item \textbf{retrieveRelevantHeaders}: Returns a string containing headers that may be relevant at the cursor location
    \item \textbf{errorReport}: Returns an error report that can be used to determine if an error round is needed, and if so, how the feedback should be presented to the LLM.
\end{enumerate}
\vspace{3pt}


This API gives leeway to the language server to decide how to implement these commands. For a language with a rich static analyzer, e.g. GHC (Haskell) with its support for hole-oriented programming and existing functionality to retrieve relevant headers (e.g. see the work of Gissurarson\cite{gissurarson2022hole}), it should be very straightforward to implement these five ChatLSP-specific commands.

To sketch the language server side of this interface, we collect the \autoref{sec:semantic-retrieval} static contextualization API below. First, we define the following types aliases:

\vspace{3pt}
\begin{itemize}
    \item \li{type} \codepat{Header} \li{=} \codetype{(Name, Type)}
    \item \li{type} \codepat{Context} \li{=} \codetype{[Header]}  
\end{itemize}

\subsection{Static Contextualization Language Server API}

\begin{itemize}
    \item \li{getExpectedType:} \codetype{(Program, LexicalLocation) -> Type}
    \item \li{getTypingContext:} \codetype{(Program, LexicalLocation) -> Context}
    \item \li{extractAliases:} \codetype{Type -> [TypeAlias]}
    \item \li{getTypeDefinition:} \codetype{TypeAlias -> Type}
    \item \li{getTargetTypes:} \codetype{Type -> [Type]}
    \item \li{filterContext:} \codetype{Context, Type -> Context}
    \item \li{scoreEntry:} \codetype{Header -> Float}
    \item \li{getStaticErrors:} \codetype{Program -> [StaticError]}
\end{itemize}

ChatLSP API methods (2) and (5)  correspond directly to \li{getExpectedType} and \li{getStaticErrors}. The following pseudocode outlines how methods (3) and (4) could be implemented using the Static Contextualization Language Server API:


\begin{lstlisting}[style=haskellpseudo]
retrieveRelevantTypes :: Type -> [Type]
retrieveRelevantTypes t = concatMap (\alias -> 
  let def = getTypeDefinition alias
  in def : getRelevantTypes def) (extractAliases t)

retrieveRelevantHeaders :: Type -> Context -> [Header]
retrieveRelevantHeaders t context = 
  let relevantTypes = retrieveRelevantTypes t
      filteredHeaders = concatMap (filterContext context) relevantTypes
      sortedHeaders = sortBy scoreEntry filteredHeaders
   in take NUMHEADERS sortedHeaders

-- Usage (given a Program and a LexicalLocation)
retrieveRelevantTypes(getExpectedType(Program, LexicalLocation))
relevantHeaders = retrieveRelevantHeaders 
  (getExpectedType Program LexicalLocation)
  (getTypingContext Program LexicalLocation)


\end{lstlisting}








\section{Related work}
\label{sec:related-work}

The introduction covered the broader literature on LLMs for code, so we focus here specifically on other methods for semantic contextualization of LLM-based code generation systems. 

Error correction using instruction-tuned models is a widespread technique and not itself a novel contribution of this paper, e.g. much work on program repair with LLMs is fundamentally rooted in this idea~\cite{prenner2021automatic,joshi2022repair}. The contribution of this paper is the observation that error looping alone is not sufficient in a context-poor setting, and that error looping together with contextualization is the most effective technique, particularly for a low-resource language like Hazel.

The observation that LLMs perform poorly when they lack repository-level context has been made in a number of recent papers, which have approached it in a variety of ways. We discussed the benchmarks used in these papers in \autoref{sec:tasks} so we do not repeat the discussion here.

RepoCoder~\cite{zhang2023repocoder} uses vector retrieval to contextualize Python code. Our experiments demonstrate that vector retrieval is sensitive to semantic confounds easily handled by static retrieval.

The Repo-Level Prompt Generator~\cite{ShrivastavaRepo} uses machine learning to decide how to construct a useful prompt, drawing information from coarse-grained static information like imports and parent-child relationships between classes. Even this level of contextualization showed substantial promise relative to baselines.

\citet{pei2023better} tackle the difficult problem of contextualizing Python function calls using a static analyzer from Python, which can provide function implementations and function usage examples. Again, even this level of contextualization is quite helpful. Our focus here was on gradually typed languages, and we did not include function implementations or usage examples, suggesting potential future work.

CodeTrek \cite{pashakhanloo2021codetrek} also uses program analyses, generated from task- and error-relevant queries and expressed in CodeQL, to generate semantic contextualization for program repair tasks in Python. This too was quite effective and suggests that richer static analyses might be of interest in particular settings. For hole filling, however, it may be that lightweight static methods, like type checking, are more efficient. However, we look forward to future direct comparisons of these methods.

\citet{li2024enhancing} also identify the semantic contextualization problem and propose IDECoder, a system that uses the static information tracked by an IDE or language server to contextualize LLM code completion. This is an outline of early experiments in this direction which have not yet been fully evaluated, but we agree with the thrusts of the argument made here and look forward to additional experimentation in this direction by the community.

CoCoMIC~\cite{ding2023cocomic} is a framework that learns in-file and cross-file context jointly atop an LLM. This differs from our approach in that it  deploys a learning step to decide which cross-file context to attend to, which may be subject to similar issues as vector retrieval approaches when given confounding contexts. However, this represents a fascinating future direction when combined with static retrieval, which can often lead to too much information to include in a token window.

In a similar vein, RLCoder~\cite{wang2024rlcoder} uses reinforcement learning to rank retrieved code snippets for repository-level code completion. Seemingly uniquely, they do not simply return the top k candidates, but impose a stop threshold, which may result in no candidates being added to the prompt if they are deemed of negative worth. Our StarCoder results suggest smaller models are especially sensitive to plausible but irrelevant inclusions, further supporting this line of investigation. A similar RL-based approach using statically derived candidates seems a promising future direction.

Dehallucinator~\cite{eghbali2024dehallucinator} is an approach that performs semantic lookup after an initial generation phase to lookup potentially relevant definitions that were invalid, e.g. not in scope. This is a more sophisticated form of error correction and could be combined with the kind of proactive contextualization that we've described. 

\citet{agrawal2023guiding} and \citet{CopilotingTheCopilots} propose an approach that modifies token sampling by leveraging the semantic code completion systems already available in modern IDEs, which implicitly provide some context. One issue with this approach is that they can only sample from tokens that the model has assigned some baseline level of probability, but without semantic context this may not be the case. 
There is likely substantial room for future work in combining static retrieval with this sort of structure-guided sampling, and perhaps with 
providing more fine-grained retrieval at each token rather than once at the onset of code completion.

\citet{zan2023private} retrieves potentially relevant code from API documentation, then further proposes a continuous training approach to incorporate this information into the model weights. In contrast, our approach is focused on black-box pre-trained models. In the future, incorporating commonly used private APIs into a continuous training loop would improve token efficiency, leaving more room in the context for truly novel definitions.

\citet{zhang2024autocoderover} builds a prompt context for program repair by retrieving class signatures and method implementations based on model-extracted keywords from GitHub issue descriptions. This is similar (and likely complementary) to our approach in that the authors define an LSP-like API for retrieval, but base this retrieval on inferring intent from unstructured text rather than cursor-local derived semantics.

\citet{chakraborty2024neuralsynthesissmtassistedprooforiented} uses RAG-based methods to retrieve relevant types from a large corpus to support synthesis of programs/proofs in the dependently-typed F* language. Corpus-based RAG on types is complementary to our approach, as it provides an avenue to retrieve semantically-similar code in cases where there are no local values with appropriate types.

\citet{parasaram2024factselectionproblemllmbased} examines the issue of `fact selection' for program repair prompt construction: How to decide which context to include and how that decision effects performance. They consider multiple types of static and dynamic information, with particular focus on localized dynamics, complementary to our more specific treatment of localized static information.

\citet{STALL} have very recently proposed a general framework for applying static analysis to repository-level code completion. They consider integration across three phases: prompting, decoding, and post-processing, the first and last corresponding to our static contextualization and correction approaches. In particular, their `token-level dependency analysis', which uses Java/Python static analyzers to add a list of plausible next tokens to the prompt, is similar to our header retrieval strategy.

Significant industry work in contextualizing code generation includes the now-standard keyword and vector embeddings approaches (as used for example by Sourcegraph Cody \cite{codyContext2023}), but many recognize new approaches are needed: the authors of the Cursor AI Code Editor call for better multi-hop retrieval \cite{cursorProblems2024}, a natural fit for structured scope-and-semantics-aware contextualization. The Zed editor features affordances for programmers to manually build and inspect prompt contexts \cite{zedAI2024} which may facilitate exploration of the relative benefits of different contextualization methods. The Aider `AI pair programmer' uses Tree-sitter ASTs to augment prompts with a condensed whole-codebase map \cite{aiderMap2023}, an approach we believe may synergize with using cursor-local semantic information to control the granularity of such a projection.

Finally, we note that there is also a vast literature on non-LLM-based code generation systems, some of which also use types to restrict the search space~\cite{Myth}. Our approach helps bring these two worlds together, e.g. by using a form of typed term enumeration to generate the relevant headers. We hope that our results will lead to more interactions between the programming languages and the AI communities.

\section{Discussion and Conclusion}
An AI model, no matter how powerful, cannot determine a human's intent 
without access to necessary context. Most existing attempts to provide this contextualization are lexically grounded, deriving from loose, associative methods developed for natural language. We believe that typed holes provide a bridge between local expressions of human intent and broader semantic context, and that type theory
provides a formal characterization of contextualization,
rooted fundamentally in the notion of \emph{typing contexts}.
In particular, {contextual modal type theory} (CMTT)~\cite{nanevski2008contextual} and gradual type theory~\cite{DBLP:conf/snapl/SiekVCB15} provide a foundation for program
sketching with holes, where expression holes corresponding to metavariables with a
corresponding type and typing context and type holes correspond to unknown types. The Hazel programming environment, with its roots in gradual CMTT~\cite{HazelnutLive} as a foundational theory of holes and its support for total syntax and type error recovery with holes~\cite{tylr2023,hazel-popl24}, therefore presents 
an ideal environment for \emph{statically contextualizing large language models with typed holes}. Our results demonstrate that this form of contextualization, together with some in-context prompting about the specific choices made in Hazel, a low-resource language, can take a model incapable of even basic MVU tasks up to, or nearly up to, the performance observed in a fully contextualized setting for a high resource language like
TypeScript. These ideas have been realized in a functional programming assistant, the Hazel Assistant.

These ideas can also be ported directly to other languages, like TypeScript, albeit with some difficulty due to limitations of standard language servers. 

Our comparisons to vector retrieval, a language-agnostic approach, suggests that language-aware programming assistants may significantly outperform language-agnostic retrieval systems in the short- and medium-term, and perhaps far into the future.

Additional forms of semantic contextualization, e.g. using dynamic test results passed backwards to holes~\cite{lubin2020program}, the results of various static and dynamic analyses, and the result of library searches to find helpers that may not yet be imported are interesting avenues for future work.

\section{Data Availability}

An artefact \cite{blinnStaticContextArtifact} containing the MVUBench program sketches and solutions, the raw data of our experiments, our testing harness, the source of the Hazel IDE and Language Server, and a copy of the StarCoder2 model used is available on Zenodo. The artefact is password-protected to prevent automatic scrapping of the benchmark suite; the password can be found in the artefact description. Hazel can be accessed online at \url{https://hazel.org}, with source available at \url{https://github.com/hazelgrove/hazel/}.

\section{Acknowledgements}
\label{sec:acknowledgements}

We would like to thank our referees for their helpful feedback, and our artefact reviewers for their diligent efforts.

The (public domain) programmer photo in \autoref{fig:conv-arch} and \autoref{fig:prompt-construction-1} depicts ENIAC programmer Ruth Teitelbaum \cite{ruthEniac}. The language server and language model images were prompted by the authors using the Bing AI Image Creator; Microsoft claims no copyright over generations \cite{bingImages}.

Claude and ChatGPT were used to author scripts for our testing harness and data processing pipeline, and aid in the transliteration of the MVUBench suite to TypeScript. In so doing we are putting faith in stated Anthropic policy of not including prompts in future training data \cite{anthropicPolicy}, which could be a source of data contamination as described in \autoref{sec:introduction}. After some discussion we decided that absolute purity was a losing battle, as there was already a near-certainty that the code was opened and edited in a Copilot-enabled IDE; true clean-room behavior with respect to benchmark data seems likely to prove increasingly difficult as LLM integration becomes more prevalent. 

This work was partially funded by the National Science Foundation (Award \#2238744).

\bibliographystyle{ACM-Reference-Format}
\bibliography{references}

\end{document}